%% file: main.tex
\documentclass[conference]{IEEEtran}
\IEEEoverridecommandlockouts

\input{config}

\makeatletter\newcommand{\removelatexerror}{\let\@latex@error\@gobble}\makeatother
\usepackage[backend=biber,maxbibnames=6,maxcitenames=6]{biblatex}
\def\BibTeX{{\rm B\kern-.05em{\sc i\kern-.025em b}\kern-.08em
		T\kern-.1667em\lower.7ex\hbox{E}\kern-.125emX}}

\begin{document}

\title{Quantum Adiabatic Generation\\of Human-Like Passwords}

\author{
\IEEEauthorblockN{\hspace*{1cm}}\and
\IEEEauthorblockN{Sascha M\"ucke\IEEEauthorrefmark{1}\IEEEauthorrefmark{5}}
\IEEEauthorblockA{sascha.muecke@tu-dortmund.de}
\and
\IEEEauthorblockN{Raoul Heese\IEEEauthorrefmark{4}}
\IEEEauthorblockA{raoul.heese@nttdata.com}
\and
\IEEEauthorblockN{Thore Gerlach\IEEEauthorrefmark{2}\IEEEauthorrefmark{5}}
\IEEEauthorblockA{gerlach@iai.uni-bonn.de}
\and
\IEEEauthorblockN{\hspace*{2cm}}\and
\IEEEauthorblockN{David	Biesner\IEEEauthorrefmark{3}\IEEEauthorrefmark{5}}
\IEEEauthorblockA{david.biesner@iais.fraunhofer.de}
\and
\IEEEauthorblockN{Loong Kuan Lee\IEEEauthorrefmark{3}\IEEEauthorrefmark{5}}
\IEEEauthorblockA{loong.kuan.lee@iais.fraunhofer.de}
\and
\IEEEauthorblockN{Nico Piatkowski\IEEEauthorrefmark{3}\IEEEauthorrefmark{5}}
\IEEEauthorblockA{nico.piatkowski@iais.fraunhofer.de}
\and
\IEEEauthorblockA{\hspace*{1.75cm}%
	\IEEEauthorrefmark{1}TU Dortmund University, %
   	\IEEEauthorrefmark{2}University of Bonn, %
   	\IEEEauthorrefmark{3}Fraunhofer IAIS, %
   	\IEEEauthorrefmark{4}NTT Data, %
   	\IEEEauthorrefmark{5}Lamarr Institute%
   	}
}


\maketitle

\begin{abstract}
    \input{abstract}
\end{abstract}


\glsresetall
\input{body}

\section*{Acknowledgment}
This research has been partially funded by the Federal Ministry of Education and Research of Germany and the state of North Rhine-Westphalia as part of the Lamarr Institute for Machine Learning and Artificial Intelligence.

\printbibliography
\end{document}

%% file: config.tex
\usepackage{amsmath,amssymb,amsfonts}
\usepackage{algorithmic}
\usepackage{graphicx}
\usepackage{textcomp}
\usepackage{xcolor}
\usepackage{tikz}
\usetikzlibrary{positioning,arrows,arrows.meta}
\usepackage{pgfplots}
\usepackage{pgfplotstable}
\pgfplotsset{compat=newest}
\usetikzlibrary{positioning}
\usepackage[ruled,norelsize]{algorithm2e}
\usepackage{bm}
\usepackage{booktabs}
\usepackage[inline]{enumitem}
\usepackage{mathtools}
\usepackage{xifthen}
\usepackage{xspace}
\usepackage{braket}
\usepackage{siunitx}
\usepackage{csquotes}
\usepackage[acronym]{glossaries}
\usepackage[hidelinks]{hyperref}
\usepackage[capitalise]{cleveref}
\usepackage{nicefrac}
\usepackage{xspace}

\newtheorem{definition}{Definition}
\Crefname{definition}{Def.}{Defs.}%

\newcommand{\mP}{\mathbb{P}}

\newcommand{\mR}{\mathbb{R}}

\newcommand{\bx}{\bm{x}}

\newcommand{\cD}{\mathcal{D}}

\newcommand{\bQ}{\bm{Q}}
\newcommand{\bU}{\bm{U}}
\newcommand{\bw}{\bm{w}}
\newcommand{\bz}{\bm{z}}
\newcommand{\bZ}{\bm{Z}}

\newcommand{\BN}{\mathbb{N}}
\newcommand{\BR}{\mathbb{R}}
\newcommand{\BP}{\mathbb{P}}

\newcommand{\T}{\intercal}
\newcommand{\BB}{\mathbb{B}}
\newcommand{\BE}{\mathbb{E}}
\newcommand{\cT}{\mathcal{T}}
\newcommand{\cZ}{\mathcal{Z}}
\newcommand{\alphabet}{\Sigma}
\newcommand{\tokenset}{\mathcal{T}}

\newcommand{\unitvec}[2][]{\ifthenelse{\isempty{#1}}{\bm{e}}{\bm{e}^{#1}}_{#2}}
\newcommand{\bin}[2][]{\ifthenelse{\isempty{#1}}{\operatorname{Bin}}{\operatorname{Bin}_{#1}}(#2)} 

\DeclareMathOperator{\KLDiv}{D_{\mathrm{KL}}}
\DeclareMathOperator{\EditDist}{ED}
\DeclareMathOperator{\MinEditDist}{MED}
\DeclarePairedDelimiter\abs\lvert\rvert
\DeclarePairedDelimiter\ceil\lceil\rceil

\DeclarePairedDelimiter\norm\lVert\rVert

\newcommand{\ie}{i.\,e.\xspace}
\newcommand{\eg}{e.\,g.\xspace}
\newcommand{\EOSToken}{\texttt{\#}\xspace}
\newcommand{\Aquila}{\emph{Aquila}\xspace}
\newcommand{\RockYou}{RockYou\xspace}

\newacronym{GenAI}{GenAI}{Generative Artificial Intelligence}
\newacronym{NLP}{NLP}{Natural Language Processing}
\newacronym{LLM}{LLM}{Large Language Models}
\newacronym{MCMC}{MCMC}{Markov Chain Monte Carlo}
\newacronym{QUBO}{QUBO}{Quadratic Unconstrained Binary Optimization}
\newacronym{UD-MIS}{UD-MIS}{Unit-Disk Maximum Independent Set}
\newacronym{MIS}{MIS}{Maximum Independent Set}
\newacronym{QC}{QC}{Quantum Computing}
\newacronym{AQC}{AQC}{Adiabatic Quantum Computing}

%% file: abstract.tex
\Gls{GenAI} for \gls{NLP} is the predominant AI technology to date. 
An important perspective for \gls{QC} is the question whether \gls{QC} has the potential to 
reduce the vast resource requirements for training and operating GenAI models. 
While large-scale generative NLP tasks are currently out of reach for practical quantum computers, 
the generation of short semantic structures such as passwords is not. 
Generating passwords that mimic real user behavior has many applications, for example to test an authentication system against realistic threat models.
Classical password generation via deep learning have recently been investigated with significant
progress in their ability to generate novel, realistic password candidates. 
In the present work we investigate the utility of adiabatic quantum computers for this task. 
More precisely, we study different encodings of token strings and propose 
novel approaches based on 
the \gls{QUBO}
and the \gls{UD-MIS} problems. 
Our approach allows us to estimate the token distribution from data and adiabatically prepare a quantum state from which we eventually sample the generated passwords via measurements. 
Our results show that relatively small samples of 128 passwords, generated on the QuEra \Aquila 256-qubit neutral atom quantum computer, 
contain human-like passwords such as \texttt{Tunas200992} or \texttt{teedem28iglove}.

%% file: body.tex
\section{Introduction}

\Glspl{LLM}, based on transformer networks, have emerged as a disruptive milestone in the recent history of artificial intelligence. The transformer architecture \cite{VaswaniSPUJGKP17} with its self-attention mechanisms, has facilitated the training of \glspl{LLM} on vast datasets, resulting in significant advancements in tasks such as translation, summarization, and conversational agents. 
Training \glspl{LLM} requires substantial computational resources, leading to significant energy consumption. 
Accurately estimating the carbon footprint associated with \glspl{LLM} even before their training 
is a significant concern and subject to ongoing research \cite{FaizKWOS0024}. 
One way to reduce the resource consumption is the utilization of more efficient algorithms \cite{deepseekai2025deepseekv3technicalreport}, 
however, another approach could be the transition to a different hardware platform. 
Thus, given recent advanced in \gls{QC}, it is reasonable to ask whether \gls{QC} has the potential to 
reduce the vast resource requirements for training und operating \gls{GenAI} models. 
Practical \gls{QC} is still in its infancy when it comes to system size and reliability, 
which renders them unusable for realizing state-of-the-art language models. 
Nevertheless, it turns out that certain real-world generative natural language models work well with 
rather moderate model sizes---one of them being password generation, also known as \textit{password guessing} \cite{BiesnerCGSK21}.

With the increasing digitalization and intercommunication between different IT systems, password security is an important cornerstone of data protection against unauthorized access and privacy violations. Therefore, an important research question is what typical users' passwords look like and whether there are common features or similarities that passwords of a group of users have in common. The existence of such patterns would imply that users from this group do not choose their passwords completely at random, but with some statistically measurable bias. This is not surprising, since users may prefer passwords that are easy to remember, for example based on language, pop culture, or personal information, which often leads to predictable structures. In addition, passwords must be typed, which can lead to easy-to-type strings being preferred, \eg, by avoiding characters that are more difficult to reach on a particular keyboard layout. Knowing the structural similarities of human-generated passwords and being able to generate similar new passwords can be very useful, for example, to test an authentication system against real-world threats, to create believable honeywords for increased security, or to evaluate the quality of new passwords with the goal of encouraging users to choose strings that avoid common patterns.

Due to the relevance of these topics, there is already a wide variety of approaches for password generators that are able mimic the patterns of training data in form of real-world password lists; see, for example, \cite{BiesnerCS22} and references therein.

In this work, we take a new perspective on the topic by combining password generation with quantum computing. Because of the increasing performance and availability of quantum computers, their use is becoming more and more relevant for practical applications. Quantum computers behave in an inherently non-deterministic way, a property that is based on quantum mechanical principles. This inherent randomness makes them very promising as generative models, especially for use cases where true randomness might be of interest for security reasons.

So far, quantum password generation remains a virtually unexplored topic. While quantum-based random number generation is a well-established technology \cite{Mannalatha2023,HeeseWMFP24}, utilizing this randomness specifically for human-like password generation is a novel use case. Interestingly, there are parallels between password generation and \gls{NLP}. Most importantly, human-generated passwords often reflect linguistic structures, incorporating dictionary words, phonemes, or even grammatical patterns. In addition, many password generation techniques use probabilistic models similar to \gls{GenAI} that predict character sequences based on past patterns. However, while passwords may contain dictionary words, they don't necessarily follow grammatical rules, and they often contain numbers or symbols in ``unnatural'' ways. Most relevant to today's resource-constrained quantum computers is the fact that passwords are typically limited by length and character diversity. As a result, password structures are typically much less complex than natural language.

\begin{figure}[t!]
	\centering
\begin{tikzpicture}[
module/.style={draw, very thick, rounded corners, minimum width=4.5cm, outer sep=0pt},
embmodule/.style={module, fill=yellow!20},
mhamodule/.style={module, fill=yellow!20},
lnmodule/.style={module, fill=yellow!20},
ffnmodule/.style={module, fill=yellow!20},
arrow/.style={-stealth', thick, rounded corners, color=black}
]
\node[embmodule] (l1) {Password Data \cite{rockyou2021}};
\node[above of= l1, mhamodule, align=center] (l2) {Learn $\mathbb{P}_{\bQ}(\bx)$};
\node[above of= l2, lnmodule, align=center] (l3) {Adiabatic evolution via $\bQ$\\or induced \gls{UD-MIS}};
\node[above of= l3, ffnmodule, align=center] (l4) {$\ket{\psi} \equiv$ password dist.};
\node[above of= l4, lnmodule, align=center] (l5) {Sample passwords\\via measurements};
\node[above of= l5] (l6) {\texttt{Tunas200992}, \texttt{teedem28iglove}, ...};

\foreach \i/\j in {l1/l2, l2/l3, l3/l4, l4/l5, l5/l6} {
    \draw[arrow] (\i) -- (\j);
}
\end{tikzpicture}
\caption{Based on historical password data, we learn the Boltzmann distribution $\mathbb{P}_{\bQ}(\bx)$ parametrized via $\bQ$. The \gls{QUBO} matrix $\bQ$ is either used directly or via our novel \gls{UD-MIS} construction to induce a quantum state $\ket{\psi}$ that encodes a probability distribution over passwords. Every measurement of $\ket{\psi}$ results in a newly generated password. The exemplary passwords shown on top are generated with the QuEra \Aquila 256-qubit neutral atom quantum computer.}
\label{fig:intro}
\end{figure}
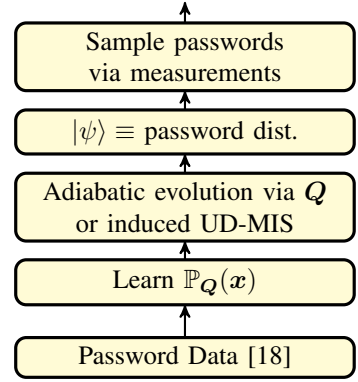

Adiabatic quantum computers, which enable quantum computation via an adiabatic state evolution \cite{albash2018}, enjoy relatively large system sizes and exhibit relatively reliable output behavior, which makes them suitable for various tasks \cite{Henriet2020,wurtz2023aquilaqueras256qubitneutralatom,Kingetal25a}. 
We will explain how to learn a probabilistic model from real-world password data \cite{rockyou2021} and 
sample passwords from two adiabatic compute architectures, as sketched in \cref{fig:intro}. 
We will not focus on how to obtain consistent samples from a desired distribution via quantum annealing---we refer to \cite{pochart.etal.2022a} as well as \cite{sato.etal.2021a,shibukawa.etal.2024a} for an elaborate discussion of this topic. Our contributions include the 
encoding of token sequences, learning of a variational password distribution parametrized via a \gls{QUBO}, and converting the \gls{QUBO} into a \gls{UD-MIS} problem by re-interpreting force-directed graph drawing as an atom placement problem. 
\footnote{The class of \gls{UD-MIS} problems is a sub-class of \gls{MIS} problems, restricted to unit disk graphs.}
Our study is wrapped up by presenting passwords generated from 
the ideal Boltzmann distribution via Markov Chain Monte Carlo, and passwords generated with the QuEra \Aquila 256-qubit neutral atom quantum computer~\cite{wurtz2023aquilaqueras256qubitneutralatom}.

\section{Background}
To begin with, we briefly summarize the basics of data-based password generation. Moreover, since different quantum hardware architectures lead to different computational paradigms, we introduce \gls{QUBO} and \gls{UD-MIS} problems. 

\subsection{Password Generation}

Let $\alphabet$ be a finite alphabet (typically consisting of ASCII characters).
A password $\bw_c \in \alphabet^*$ is a finite sequence of characters\footnote{A star superscript is used to denote the set of all finite strings over an alphabet.} from $\alphabet$.
For technical and practical reasons, passwords have a maximum length (number of characters), which we denote as $L$.
In addition, we also presume a minimum password length $l \leq L$.
In the following, we consider only length-restricted passwords $\bw_c \in \alphabet^{[l,L]}$ with $l \leq \abs{\bw_c} \leq  L$, where $\alphabet^{[l,L]} := \bigcup_{k=l}^L \alphabet^k \subset \alphabet^*$ and $\alphabet^k := \{ \bw_c \mid \bw_c \in \alphabet^* \land \abs{\bw_c} = k \} \subset \alphabet^*$ represent sets of length-restricted strings.

The motivation of this work is the generation of human-like passwords based on passwords list from real-world user groups.
Formally, this password data consists of a multi-set of $N$ passwords $\cD := \{\bw_c^1,\dots,\bw_c^N\}$.
We assume that $\cD$ contains samples of a random variable $\bm{W}_c$ that follows an underlying probability distribution over the set $\alphabet^{[l,L]}$ with probability function $p_c(\bw_c)=\BP[\bm{W}_c=\bw_c]$.
This distribution is distinctly different from, \eg, a uniform distribution over all possible passwords, as human users typically prefer to choose passwords that follow certain (but not necessarily known) patterns.

\paragraph{Tokenization}
By analyzing the distribution of $n$-grams across our data set, we can define certain strings of characters from $\alphabet$ with a high co-occurrence probability, a process known as \emph{tokenization}.
This yields a new alphabet of \emph{tokens} which we can use in place of the original $\alphabet$, increasing the alphabet size while reducing the number of characters per word \cite{BiesnerCGSK21}.
In analogy to the character-based password defined above, a token $t \in \tokenset \subset \alphabet^*$ is therefore a finite sequence of characters from $\alphabet$.
Here we define the set of all tokens $\tokenset$ as a subset of the set of all finite strings over $\alphabet$ with $T:=\abs\tokenset$.
The choice of tokens is generally ambiguous.
We presume that the tokenization is performed such that all passwords from the password data $\cD$ can be represented, but not necessarily all possible words in $\alphabet^{[l,L]}$.\footnote{A way to enforce this is would be requiring $\alphabet\subseteq \tokenset$.}
To that end, we define the domain of tokenizable passwords $D_c \subseteq \alphabet^{[l,L]}$ for which we presume that $\cD \subseteq D_c$.\footnote{When a multi-set is referred to as a subset of a set, this means that the unique elements of the multi-set form a subset of the given set.}

A unique tokenization function is a mapping $\tau: D_c \rightarrow D_t$ from a password to a unique sequence of tokens of length $M$, where $M$ is the maximum length of tokens required to encode all passwords in $D_c$.
Here, $D_t := \{ \bw_t \mid \bw_t \in \tokenset^* \land |\bw_t| \leq M \} \subset \tokenset^*$ denotes the domain of feasible token sequences.
The password data $\cD$ can therefore also be represented as a set of tokenized passwords $\{ \bw_t \mid \bw_t = \tau(\bw_c) ~\forall \bw_c \in \cD \} \subseteq D_t$.
According to its definition, the tokenization function $\tau$ is to be understood as an arbitrary but fixed deterministic map that resolves any ambiguities arising from multiple possible tokenizations of the same word.
All passwords in the domain of non-tokenizable passwords $\alphabet^{[l,L]} \setminus D_c$ are treated as out-of-vocabulary (OOV) sequences.
Complementary to the tokenization $\tau$, we define the inverse tokenization $\tau^{-1}: D_t \rightarrow D_c$, which is to be understood as a string concatenation of character sequences such that $\tau^{-1}(\tau(\bw_c)) = \bw_c$ and $\tau(\tau^{-1}(\bw_t)) = \bw_t$ for all $\bw_c \in D_c$ and $\bw_t \in D_t$, respectively.
The underlying ground truth distribution of passwords can also formally be transformed into a tokenized representation, which reads 
\begin{equation}
	p_t(\bw_t) := \frac{p_c(\tau^{-1}(\bw_t))}{\sum_{\bw_c \in D_c} p_c(\bw_c) }.\label{eq:pt}
\end{equation}
By definition, the inverse tokenization only maps to $D_c \subseteq \alphabet^{[l,L]}$, which may reduce the support of $p_c$ and requires a re-normalization. Therefore, depending on the tokenization, $p_t$ may not be able to capture $p_c$ entirely.

\paragraph{Generation}
The generation of new passwords requires a model-based estimation $\hat{p}_t$ of $p_t$ based on $\cD$.
This leads to an estimate for the distribution of passwords
\begin{equation}
	\hat{p}_c(\bw_c) = \begin{cases} \hat{p}_t(\tau(\bw_c)) & \bw_c \in D_c \\ 0 & \text{otherwise} \end{cases}
\end{equation}
based on the chosen tokenization, which is defined by the tokenization function $\tau$ and the domain of tokenizable passwords $D_c$.
The goal of this work is to study exemplary how quantum computers can be used to model $\hat{p}_t$ for a given tokenization.

\subsection{Adiabatic Quantum Computing}\label{sec:aqc}
The primary goal of \gls{AQC} is to prepare a quantum system in the ground state of a problem-specific target Hamiltonian by slowly evolving an initial Hamiltonian into the target Hamiltonian under the adiabatic theorem \cite{albash2018}. More specifically, the system is initially prepared in the known ground state $\ket{\psi_0}$ of some prescribed Hamiltonian $H_0$. 
The goal is to find the ground state $\ket{\psi}$ of a problem-specific target Hamiltonian $H$ that encodes the solution to a given task.
To that end, an adiabatic (that is, sufficiently slow) evolution for $\alpha$ from $0$ to $1$ drives the system Hamiltonian $H(\alpha)=(1-\alpha)H_0 + \alpha H$ towards $H$. The construction of $H$ is directly driven by the problem at hand.

The problem classes that can be solved depend on the type of hardware that is used. For example, Ising machines with up to quadratic interaction terms allow to encode a \gls{QUBO}. A common approach is define a real symmetric matrix $\bQ$ such that, in a suitable basis, the matrix elements of the Hamiltonian are given by $(H)_{i,j} = \bx^{i \T} \bQ \bx^{j}$ where $\bx^{i}$ is the $i$-th $n$-bit binary vector in some arbitrary but fixed order. 
$\bQ \in \mathbb{R}^{n\times n}$ is the so-called \emph{\gls{QUBO} matrix}. By construction, computing the ground state $\ket{\psi}$ of $H$ is then equivalent to solving $\arg\min_{\bx\in\{0,1\}^n} \bx^{\T} \bQ \bx$.
Since \glspl{QUBO} are NP-hard \cite{PardalosJ92}, a multitude of relevant real-world problems can be reduced to them. 

The intrinsic Hamiltonian of neutral atom quantum computers gives rise to another combinatorial problem, the so-called \gls{UD-MIS} problem \cite{Ebadi2022a}. In \gls{UD-MIS}, a problem instance is specified by a graphical structure $G=(V,E)$ whereas each vertex $v\in V$ is identified with a location is a two-dimensional plane. Each vertex represents a dimension $\bx_v$ of a $n$-dimensional bit-string $\bx$. In \gls{UD-MIS}, when a vertex $v$ is on state $1$, \ie, $\bx_v=1$, all of its neighbors $N(v)=\{w : \{v,w\}\in E \}$ must take the value $0$. The goal is to choose $\bx$ such that the number of ones, \ie, $\|\bx\|_1=\sum_{i=1}^n |\bx_i|$, is maximized. 
A graph can be encoded on the quantum device by placing atoms on a two-dimensional grid such that each atom represents a node. When driving the system via an adiabatic evolution, the final state can be shown to maximize the number of atoms in the Rydberg state \cite{wurtz2023aquilaqueras256qubitneutralatom}. However, it is energetically more favorable that atoms within the so-called \emph{Rydberg blockade radius} attain a state in which only one of them is in a Rydberg state and the others are not. This is a natural representation of the \gls{MIS} constraint for a graph in which atoms within the Rydberg blockade radius are considered as neighbors. Hence, \gls{UD-MIS} problems can naturally be encoded into the Hamiltonian of such quantum systems \cite{Henriet2020}. The \gls{UD-MIS} problem is NP-complete. 

Practical quantum computers are typically not deterministic and subject to hardware-related noise. Therefore, they sample from a state that yields the solution to the respective problem instance with a certain probability (which is ideally high), but can also produce non-optimal results, introducing a certain level of uncertainty. In what follows, we show how to encode a probability mass function over token sequences into the induced \emph{Gibbs state} of a \gls{QUBO}. Finally, we explain how the \gls{QUBO} can be used to construct an \gls{UD-MIS} instance that allows us to approximately sample from the learned password distribution via neutral atom devices.

\section{Encodings and Boltzmann Distributions}

Given a fixed set of tokens $\tokenset\subset \alphabet^*$ with $T=\abs\cT$ and a tokenization function $\tau$ extracted from a text document, we can build a generative model based on the Boltzmann distribution over an Ising model or, equivalently, a \gls{QUBO} instance.
In the following, we will \begin{enumerate*}[label=(\roman*)]\item build a mapping between token strings and binary vectors, which allows us to \item construct a \gls{QUBO} instance with parameter matrix $\bQ$ whose associated Boltzmann distribution (see \cref{def:boltzmann}) matches the distribution of a given password dataset.\end{enumerate*}

\paragraph{Binary token representation}\label{sec:bintoken}
As a first step, we need to convert passwords (or, equivalently, token sequences) to binary vectors, which is the target domain of \gls{QUBO}.
To this end, we firstly assign an arbitrary but fixed order to all unique tokens and enumerate them as $\tokenset=\lbrace t^1,\dots,t^T\rbrace$.
Note that we use superscript $t^i$ to denote the $i$-th \emph{unique} token, while we use subscript $t_i$ to denote \emph{any} token with index $i$ (\ie, position $i$ in a sequence of tokens that represent a password).

Next, we need an invertible mapping $\kappa:\tokenset\rightarrow\BB^k$ from each unique token $t^i$ to a binary representation of fixed length $k$.
Clearly, we require $k\geq\log_2T$ so that all tokens can be represented.
The binary encoding of a password can then be achieved by encoding each of its tokens.
By construction, passwords can consist of up to $M$ tokens. To ensure a binary encoding of constant length, we introduce a special token \EOSToken denoting the end of a word, after which every token should be ignored.
Words with fewer than $M$ tokens are padded with \EOSToken to a length of exactly $M$ (which means that a \EOSToken token may only be followed by other \EOSToken tokens).
For the sake of simplicity, we presume that $\EOSToken\in\tokenset$.  
In conclusion, to encode a password $\bw_t \in D_t$ consisting of $\ell \leq M$ tokens, we concatenate the individual tokens' encodings and add a \EOSToken padding to fill up the remaining space: $\bw_t=$
\begin{equation*}
t_1\cdot t_2\cdot\dots\cdot t_{\ell}~\longmapsto\\~(\kappa(t_1),\kappa(t_2),\dots,\kappa(t_{\ell}),\underbrace{\kappa(\EOSToken),\dots,\kappa(\EOSToken)}_{M-l})^\T
\end{equation*}
$=:\kappa^M(\bw_t)\in\BB^{M k}$. By applying $\kappa^M$ to all token strings in our data set $\cD$ (denoting a multi-set of passwords of up to $M$ tokens), we obtain a probability distribution over binary password representations of length $Mk$ through $\BP_{\cD,\kappa}[\bZ=\bz]=$ \begin{equation}
	P(\bz;\cD,\kappa)\propto\begin{cases}
		p_t(\bw_t) &\text{if }\exists!\,\bw_t: ~\kappa^M(\bw_t)=\bz,\\ 
		0          &\text{otherwise.}
	\end{cases}
\end{equation}

There are multiple possible choices of $\kappa$, influencing the resulting probability distribution.
We shall discuss each of them briefly.
\begin{description}
	\item[One-Hot Encoding] $\kappa(t^i)=\unitvec[T]{i}$.
	While this representation allows for a unique parameter $Q_{ij}$ for every pair of tokens, it leads to very large \gls{QUBO} instances due to $k=T$.	
	\item[Binary Encoding] $\kappa(t^i)=\bin[k]{i-1}$ with $k=\ceil{\log_2T}$.
	Binary representation of the token index $i$ (counting from 0).
	This is the most compact representation in terms of the number of required binary variables.
	However, each parameter $Q_{ij}$ is shared between up to $2\log_2T$ tokens, making the resulting distribution harder to model using only $k(k+1)/2$ parameters.
	\item[Stacked Binary Encoding] $\kappa(t^i)=(\unitvec[k_1]{i_1},\dots,\unitvec[k_m]{i_m})$, such that $\forall j:~k_j\geq 2$, $\prod_{j=1}^mk_j\geq T$ and $\sum_{j=1}^m[(i_j-1)\prod_{\iota=1}^{j-1}k_{\iota}]=i$.
	A concatenation of multiple one-hot vectors, such that the total number of possible states is at least $T$.
	This is a compromise between one-hot and binary encoding, balancing the total number of bits and the number of tokens per parameter $Q_{ij}$.
\end{description}

\begin{table}
	\centering
	\caption{Binary token encodings used for experiments.
		All encodings yield exactly $256$ binary code words, $\bm{k}=(k_1,\dots,k_m)$ gives the length of individual one-hot encodings.}
	\label{tab:enc}
	\begin{tabular}{lcc}
		\toprule
		encoding &\#bits/token &$\bm{k}$ \\
		\midrule
		binary &8 &--- \\
		stacked &16 &$(2,2,2,2,2,2,2,2)$ \\
		stacked &20 &$(2,2,8,8)$ \\
		stacked &24 &$(2,2,2,2,16)$ \\
		\bottomrule
	\end{tabular}
\end{table}

\paragraph{Training a Boltzmann QUBO}
A \gls{QUBO} instance, like an Ising model, induces a \emph{Boltzmann distribution} over binary vectors through its \emph{Gibbs state}.

\begin{definition}[Boltzmann Distribution of a QUBO]\label{def:boltzmann}
Let $\bQ\in\BR^{n\times n}$ be an upper triangular matrix parametrizing a \gls{QUBO} instance with energy function $E_{\bQ}(\bz)=\bz^\T\bQ\bz$ for some fixed $n>0$.
Further, let $\beta\in\BR_+$ be an inverse temperature.
Then $\bZ$ follows a Boltzmann distribution over the set $\BB^n$ with probability function \begin{align*}
	\BP_{\bQ,\beta}[\bZ=\bz]&=P(\bz;\bQ,\beta)=Z_{\bQ,\beta}^{-1}\exp(-\beta E_{\bQ}(\bz)),\\
	\text{where } Z_{\bQ,\beta} &= \sum_{\tilde{\bz}\in\BB^n}\exp(-\beta E_{\bQ}(\tilde{\bz})).
\end{align*}
\end{definition}

The term $Z_{\bQ,\beta}$ is the \emph{partition function}, acting as a normalizing constant, and is \#P-hard to compute as it involves an explicit enumeration of all $2^n$ bit-strings. 
For $\beta\rightarrow\infty$ the Boltzmann distribution becomes degenerate, with $P(\bz^*;\bQ,\beta)\rightarrow 1$ for the minimizing vector $\bz^*$ with $E_{\bQ}(\bz^*)\leq E_{\bQ}(\tilde{\bz})~\forall\tilde{\bz}\in\BB^n$.
For $\beta\rightarrow 0$ it approaches a uniform distribution $P(\bz;\bQ,\beta)\rightarrow 2^{-n}$.
From the definition of $E_{\bQ}$, we can see that $\beta$ can be integrated into $\bQ$ by assuming that $\bQ=\beta\tilde{\bQ}$ with $\norm{\tilde{\bQ}}_{\infty}=1$, where $\norm{\cdot}_{\infty}$ denotes the max norm.
For this reason we set $\beta=1$ for the remainder of this article and drop it from our notation.

The challenge remains to find a \gls{QUBO} instance $\bQ\in\BB^{Mk\times Mk}$ whose corresponding Boltzmann distribution $\BP_{\bQ}$ approximates the target distribution $\BP_{\cD,\kappa}$ of password tokens best.
To this end, we can perform an iterative training procedure by minimizing the \emph{Kullback-Leibler divergence} between $\BP_{\bQ}$ and $\BP_{\cD,\kappa}$, which reads \begin{equation}
	\KLDiv(\BP_{\cD,\kappa}\,\Vert\,\BP_{\bQ})=\sum_{\bz\in\BB^{Mk}}P(\bz;\cD,\kappa)\log\left(\frac{P(\bz;\cD,\kappa)}{P(\bz;\bQ)}\right)
\end{equation}%
and approaches $0$ when the two distributions are identical.
The gradient of this expression w.r.t. $\bQ$ is given by \begin{align}\label{eq:grad}\begin{split}
	\bU &= \BE_{\bZ\sim\BP_{\bQ}}[\bZ\bZ^\T]-\BE_{\bZ\sim\BP_{\cD,\kappa}}[\bZ\bZ^\T]\\
		&= \sum_{\bz\in\BB^{Mk}}[P(\bz;\bQ)-P(\bz;\cD,\kappa)]\bz\bz^\T,
\end{split}\end{align}%
allowing us to repeatedly perform updates to the \gls{QUBO} parameter matrix through $\bQ\mapsto\bQ-\eta\bU$ for some learning rate $\eta>0$, until some convergence criterion is met (\eg, the KL-divergence falls below some threshold, or a budget is depleted).

Notice how \cref{eq:grad} involves the expectation of $\bZ\bZ^\T$ w.r.t. the true distribution and w.r.t. to the distribution induced from $\bQ$. 
The latter can be obtained via \gls{MCMC} or Gibbs sampling from $\BP_{\bQ}$~\cite{Hastings/1970a,Geman/Geman/1984a} while the former is 
estimated from our data set $\cD$, leading to \begin{equation}\label{eq:gradapprox}
	\hat{\bU}=\frac{1}{\abs{\cZ_{\bQ}}}\sum_{\bz\in\cZ_{\bQ}}\bz\bz^\T-\frac{1}{\abs{\cD}}\sum_{\bw_t\in \cD}\kappa^M(\bw_t)\kappa^M(\bw_t)^\T\;.
\end{equation}
Both estimators are unbiased and the expression converges to the exact gradient as both, the number of training passwords and the number of generated \gls{MCMC} samples tend to infinity. 
Due to the stochasticity of our gradient estimator, we apply ADAM \cite{KingmaB14} to mitigate the noise. 

\begin{figure}
  \removelatexerror
  \begin{algorithm}[H]
    \caption{Force-Directed Atom Placement\label{alg:forces}}
    \KwIn{\Gls{QUBO} matrix $\bQ\in\mR^{n \times n}$, vertex set $V$, stepsize $\eta$, number of iterations $I$, constant $c$}
    \KwOut{Atom locations $(v_x,v_y)$ for each $v\in V$}
    \For{$i \leftarrow 1 \to I$}{
    $v_{x} \leftarrow \cos(-2v\pi/n) / c$\;
    $v_{y} \leftarrow \sin(-2v\pi/n) / c$\;
    }
    \For{$v \in V$}{
      $\Delta_x \leftarrow 0$\;
      $\Delta_y \leftarrow 0$\;
      \For{$w \in V$}{
        \If{$v \neq w$}{
        	$d \leftarrow \|v-w\|_2$\;
        	\If{$d \leq 4 \times 10^{-6}$}{
        	  $\gamma \leftarrow \mP_{\bQ}[(v,w)\in E] (1+|N(v)|)/d$\;
        	  $\Delta_x \leftarrow \Delta_x + \gamma({v_x-w_x})/d$\;
        	  $\Delta_y \leftarrow \Delta_y + \gamma({v_y-w_y})/d$\;
	        }
        }
      }
      $v_x^{\operatorname{new}} = v_x + \eta \Delta_x$\;
      $v_y^{\operatorname{new}} = v_y + \eta \Delta_y$\;
      \If{$v^{\operatorname{new}}$ is not out of range}{
      	$(v_x,v_y) \leftarrow (v^{\operatorname{new}}_x,v^{\operatorname{new}}_y)$\;
      }
    }
    \Return $\{(v_x,v_y)\}_{v\in V}$\;
  \end{algorithm}
\end{figure}

\begin{figure}
	\centering
	\includegraphics[width=\columnwidth]{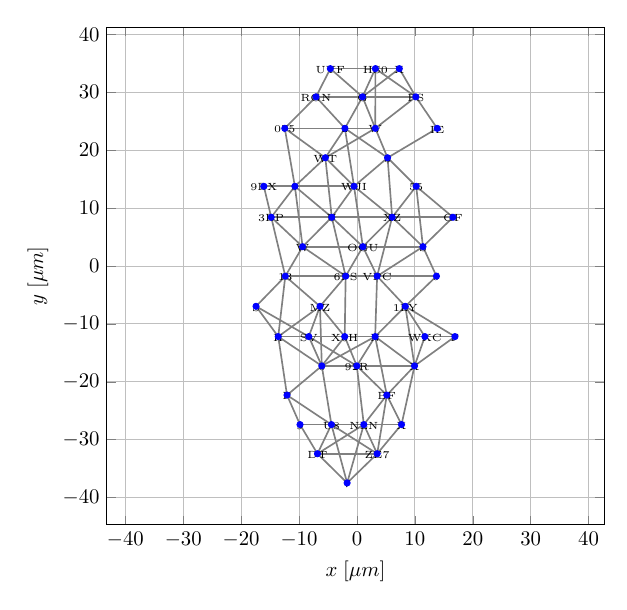}
	\caption{Exemplary graphical structure derived from the atom placement and the Rydberg blockade radius for a model with Binary(8) encoding. In this encoding, each atom corresponds to a specific token, which is annotated at each corresponding vertex.}
	\label{fig:graph}
\end{figure}

\subsection{Force-Directed Atom Placement}\label{sec:fdgd}

As explained in \cref{sec:aqc}, neutral atom quantum computers are amenable to \gls{UD-MIS} problems. We now present a novel method to 
construct \gls{UD-MIS} from our learned \gls{QUBO}. It is important to understand that our construction does not create additional variables. Instead, 
atoms repel each other based on the \gls{QUBO} weights. When $\bQ_{v,w}$ is a negative \gls{QUBO} matrix entry with a large magnitude, the marginal probability of the 
event $\bx_v=1 \wedge \bx_w=1$ is large, and vice versa.  Consequently, vertex pairs with a strongly negative $\bQ_{v,w}$-value cannot be placed close together in the \gls{UD-MIS}, since the Rydberg Blockade would prevent them from being in their \enquote{preferred} joint state $(1,1)$ (and similar for vertex pairs with strongly positive $\bQ_{v,w}$-value). Thus, we re-interpret the entries of $\bQ$ as logarithms of unnormalized \textit{edge appearance probabilities}. That is, $\mP[(v,w) \in E] = \exp(-0.5 \bQ_{v,w}) / \sum_{i} \sum_{j} \exp(-0.5 \bQ_{i,j})$, also known as soft-max of $\bQ$. 

In order to find atom locations which respect the neighborhood structure that is implied by $\mP_{\bQ}[(i,j) \in E]$, 
we consider an initial placement that is iteratively refined by applying abstract forces to the vertices. 
The procedure is inspired by the Fruchterman-Reingold graph drawing algorithm~\cite{FruchtermanR91}. 
The Fruchterman-Reingold algorithm is a force-directed graph drawing method that was originally designed to create visually appealing layouts for graphs by simulating physical forces among nodes. Normally, it operates by applying attractive forces between connected nodes, akin to springs, while simultaneously applying repulsive forces between all pairs of nodes, similar to charged particles. Through iterative adjustments of the vertex positions based on the resultant forces, the algorithm aims to minimize edge crossings and evenly distribute nodes across the drawing area, ultimately achieving a stable and clear representation of the graph structure. 

We modify the algorithm in order to obey the constraints of the physical atom placement. More precisely, the x-coordinates of any pair of vertices is are not allowed to have a distance larger than \SI{75}{\micro\meter} (x-separation), the y-coordinates of any pair of vertices is are not allowed to have a distance larger than \SI{76}{\micro\meter} (y-separation), the euclidean distance between any two points is not allowed to exceed \SI{4}{\micro\meter}, and the difference in y-coordinates of any pair of vertices is either \SI{>4}{\micro\meter} or $0$. These values are applied to define the available area for the graph embedding and to define a trigger for the repulsion. That is, repulsive forces only act on vertex pairs which are closer than \SI{4}{\micro\meter}. 

The resulting algorithm is shown in \cref{alg:forces}. As input, it expects a \gls{QUBO} matrix $\bQ$, a vertex set $V$, a real-valued stepsize $\eta$, a integer number of iterations $I$, and a real constant $c$---the latter re-scales the initial placement such that it fits on the available area. The algorithm utilizes repulsion between nearby atoms as well as rejection of new locations which would exceed the available area. Attractive forces are not required. For the sake of conciseness, we postpone a theoretical analysis of our Fruchterman-Reingold-inspired atom placement algorithm to future work. 

\begin{figure*}
	\centering
	\includegraphics[width=3.5cm]{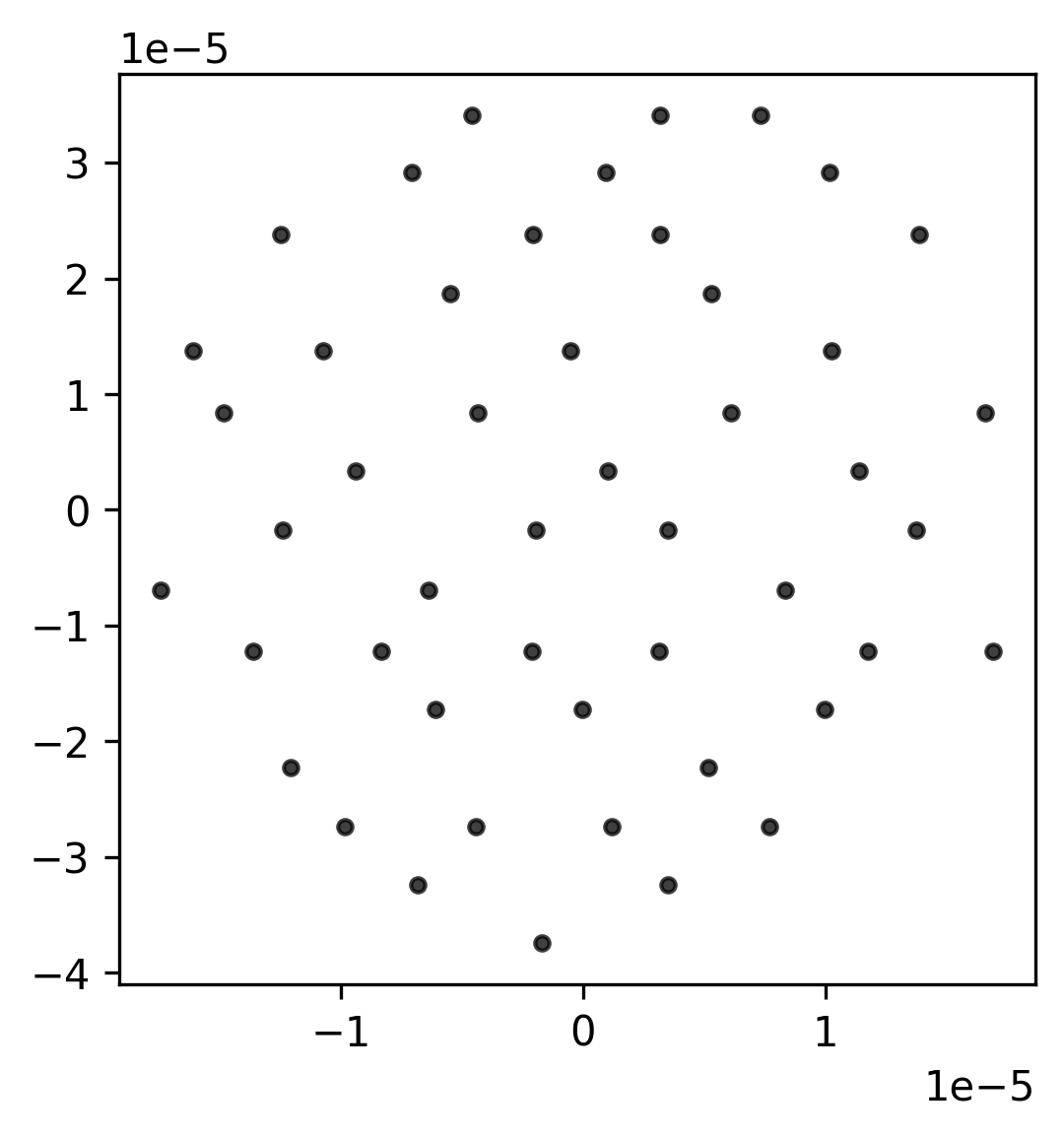}
	\includegraphics[width=3.5cm]{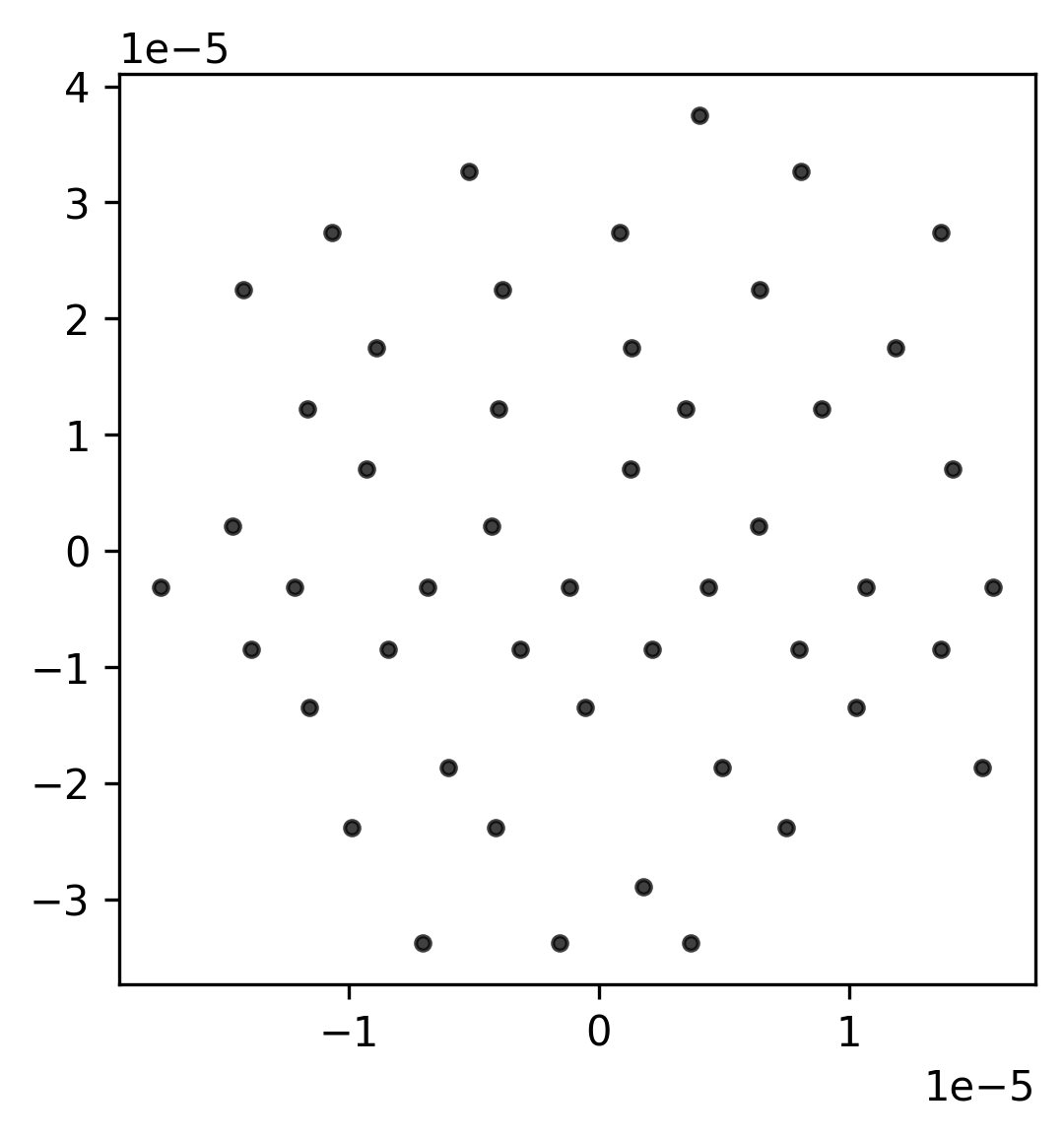}
	\includegraphics[width=3.5cm]{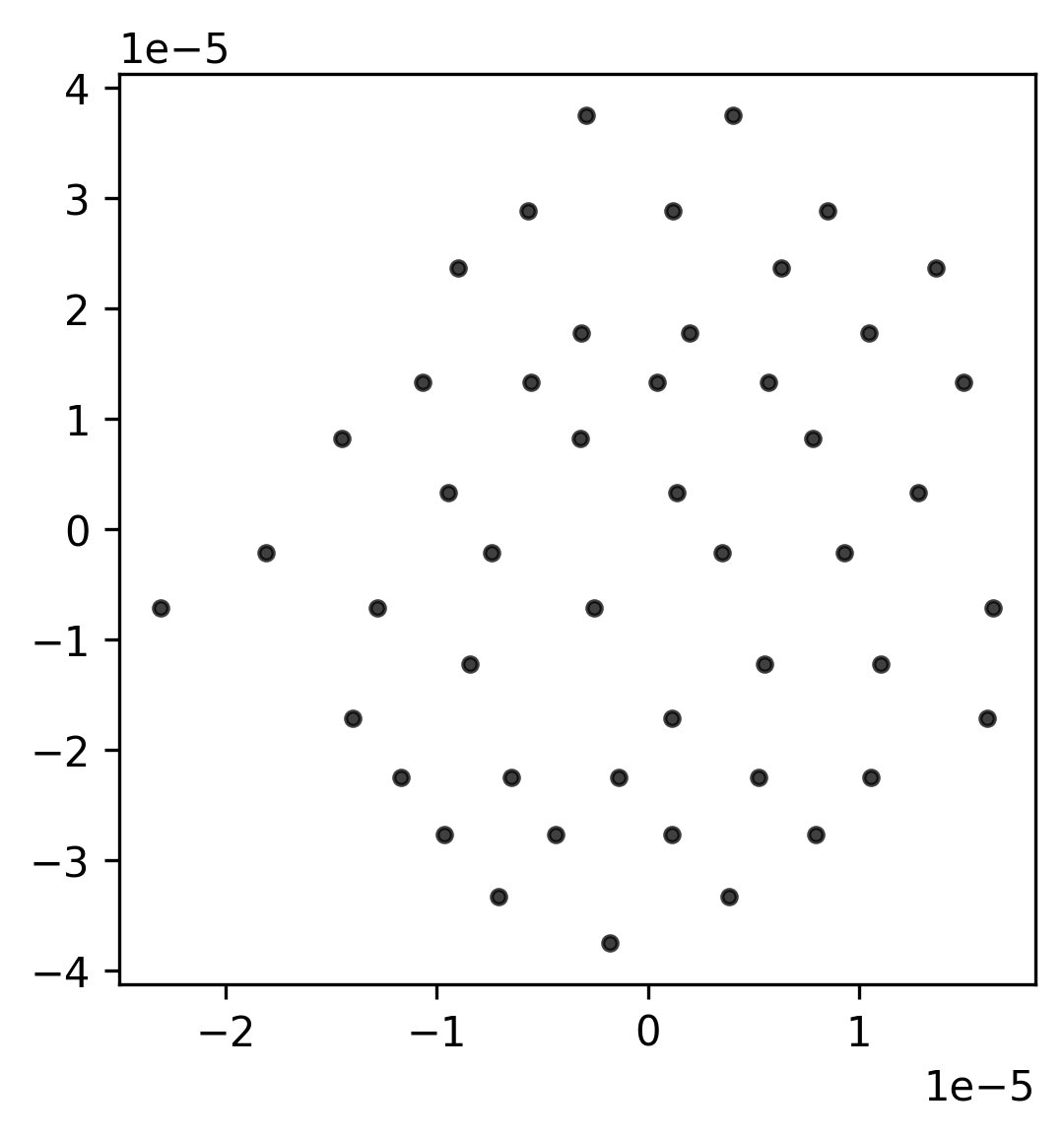}
	\includegraphics[width=3.5cm]{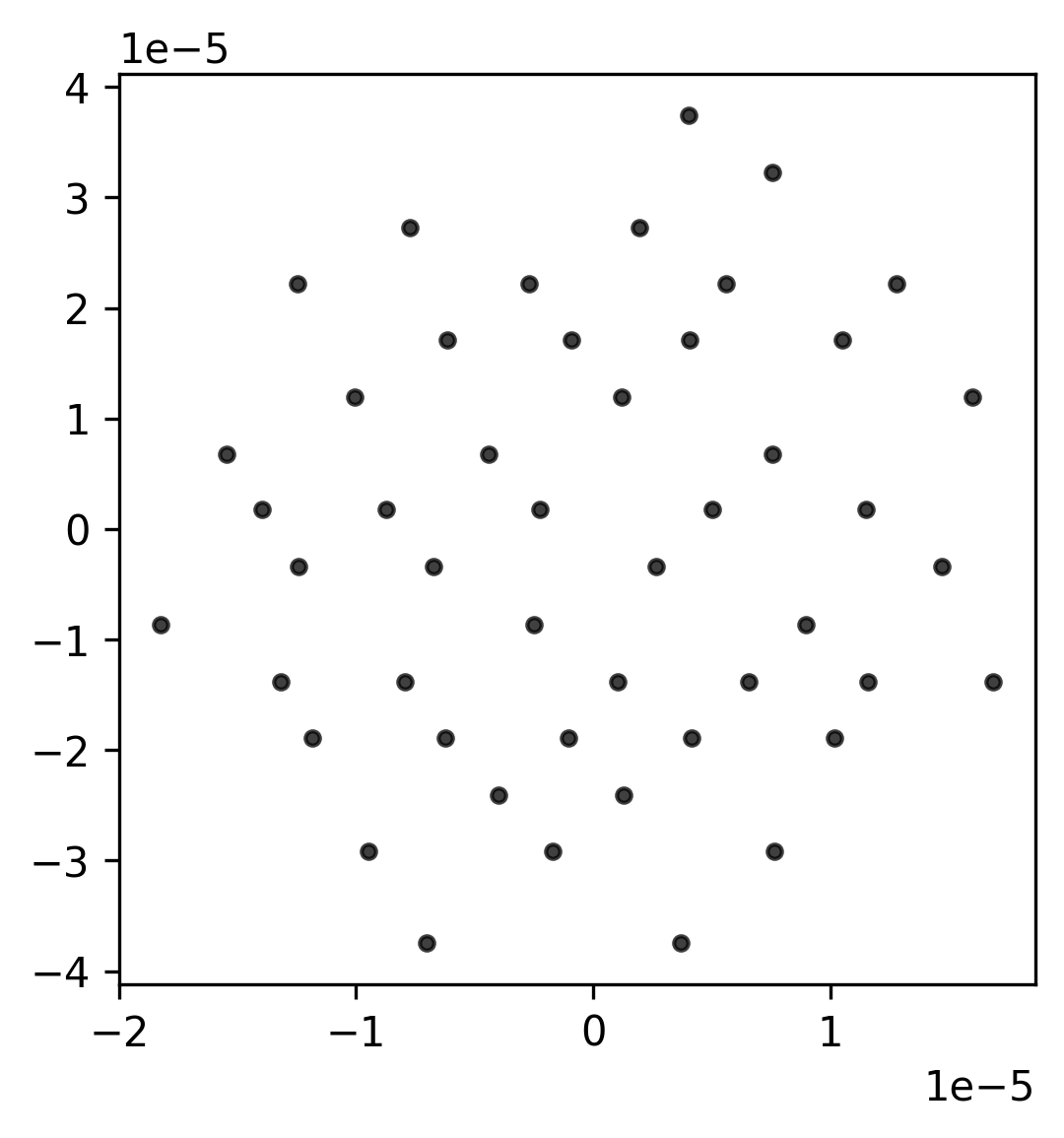}
	\includegraphics[width=3.5cm]{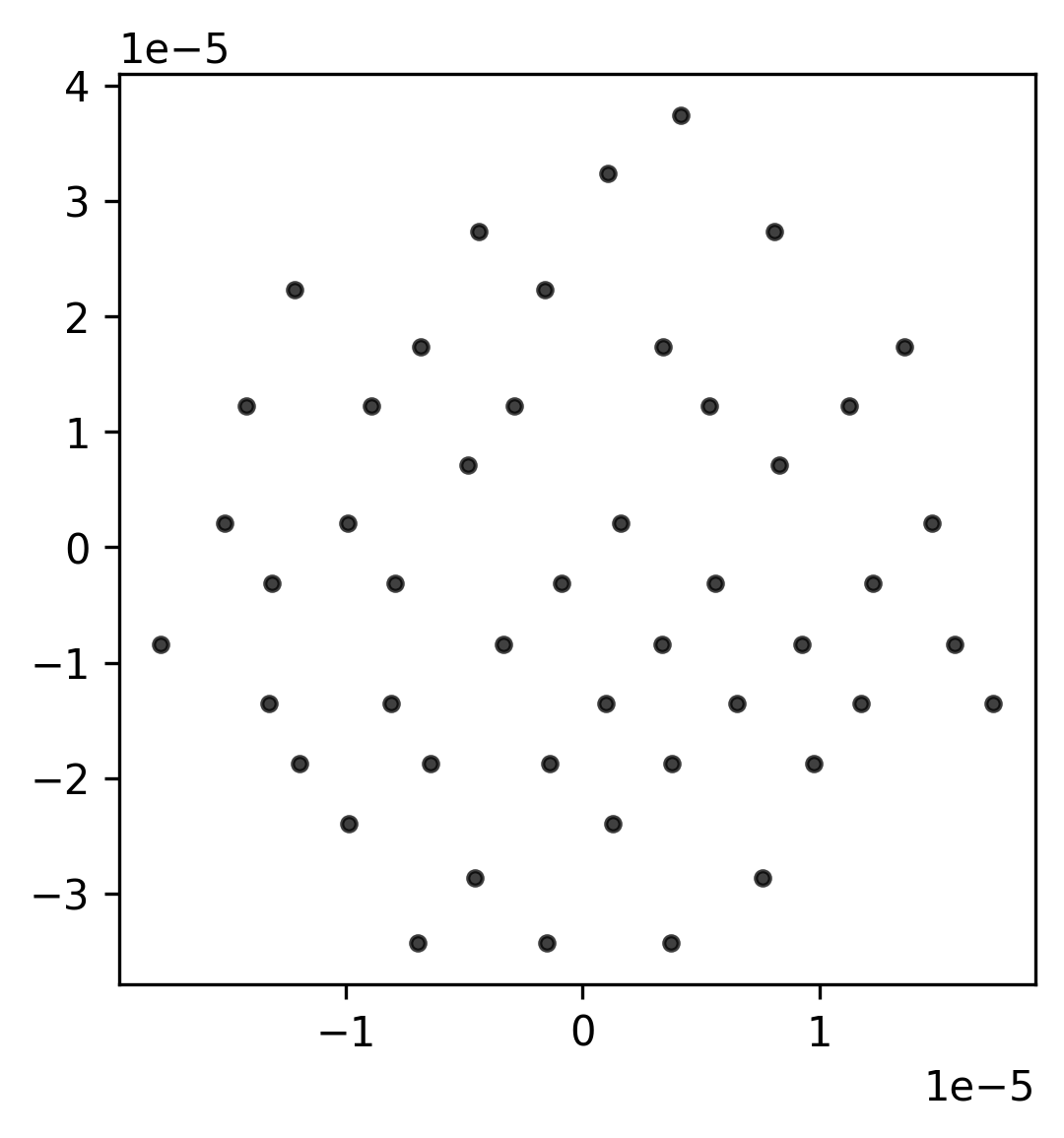}\\
	\includegraphics[width=3.5cm]{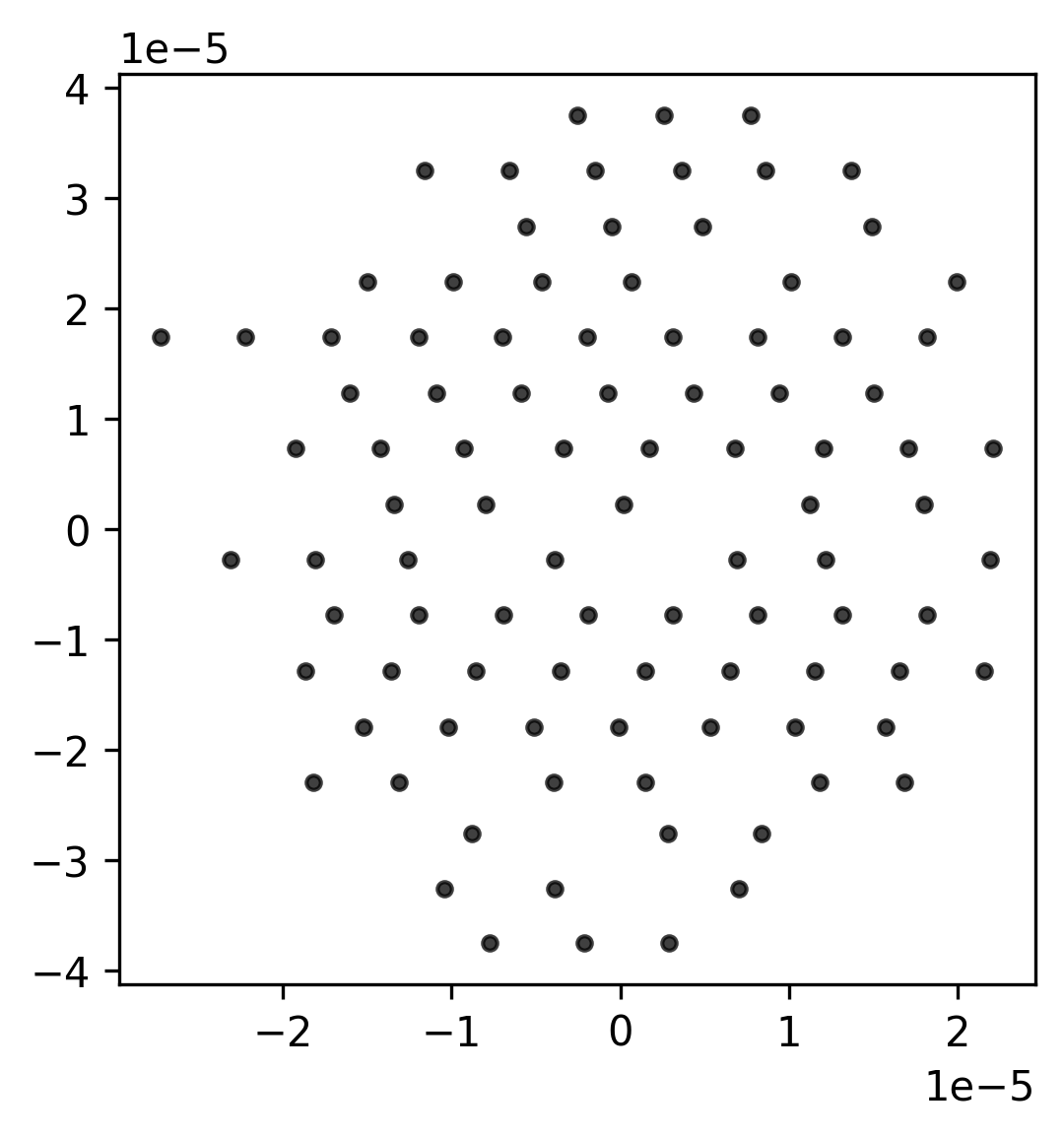}
	\includegraphics[width=3.5cm]{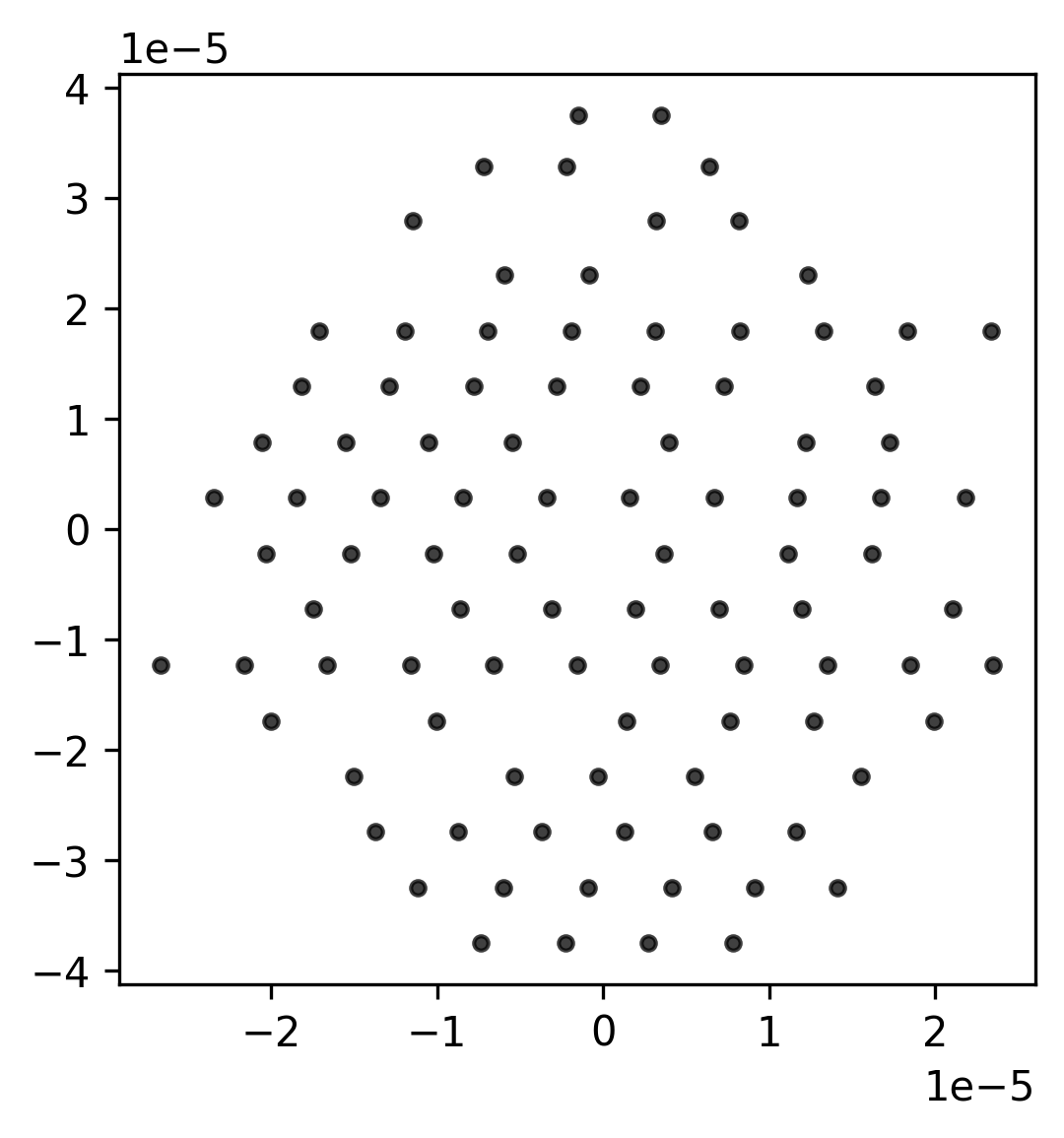}
	\includegraphics[width=3.5cm]{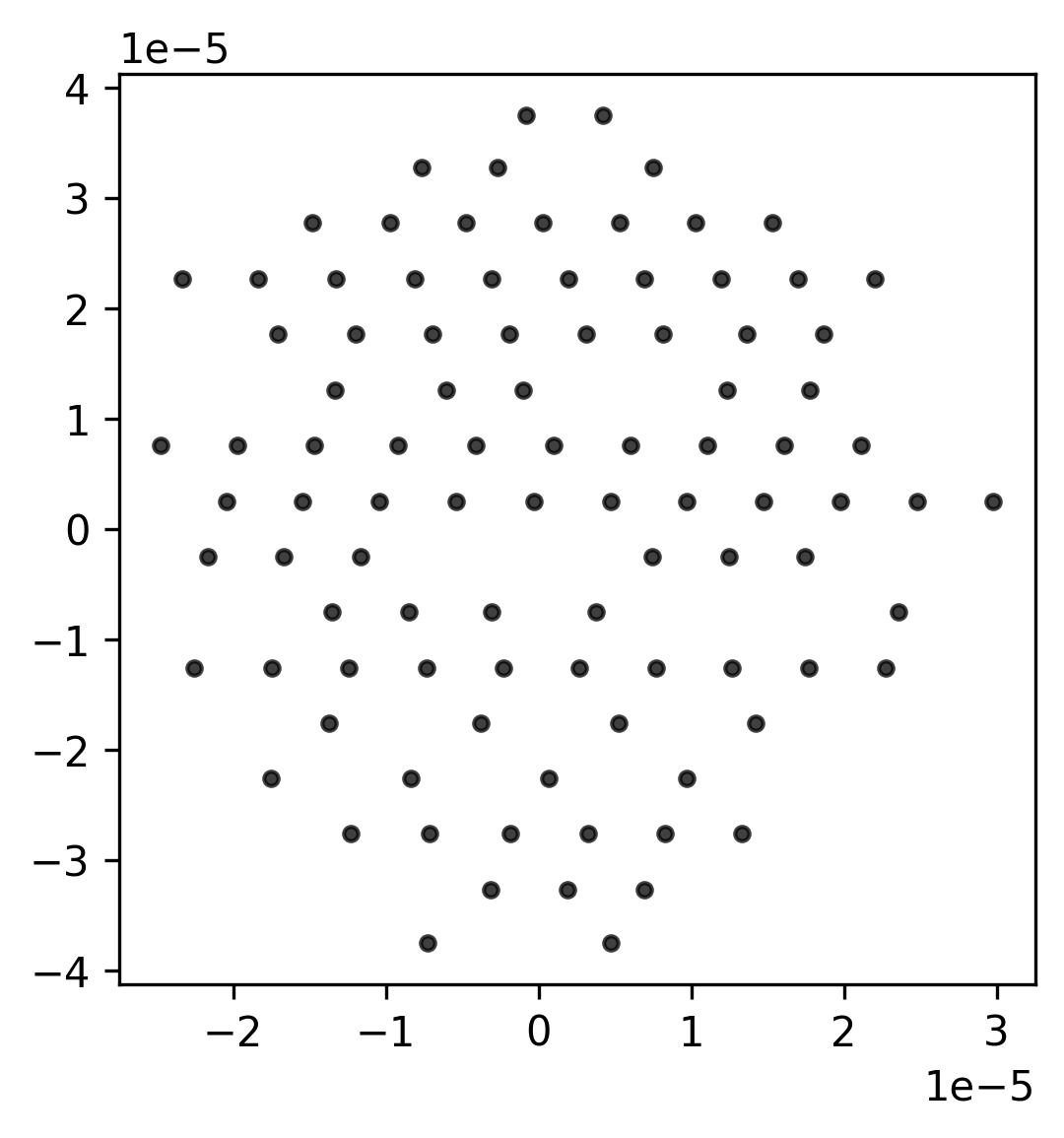}
	\includegraphics[width=3.5cm]{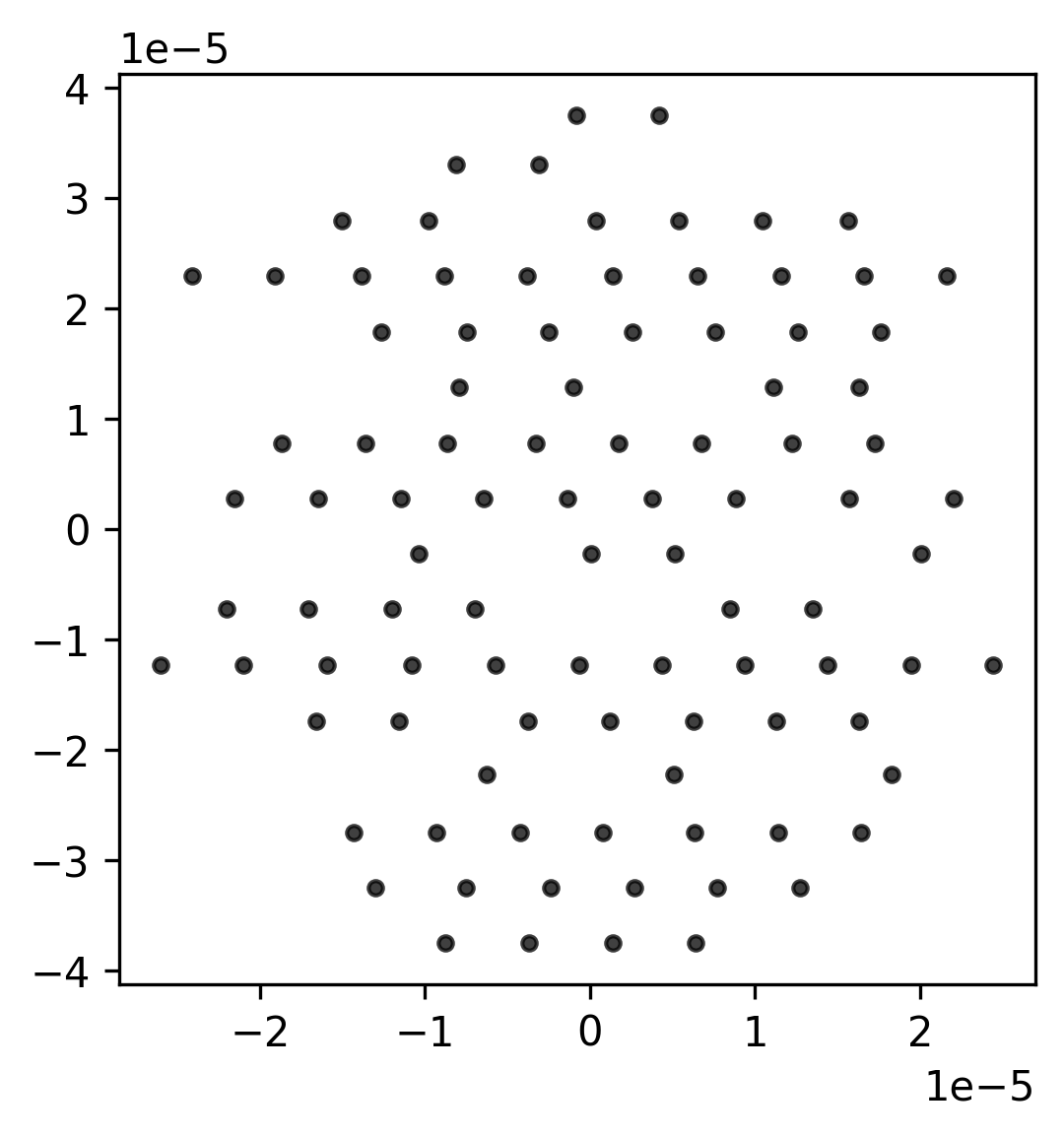}
	\includegraphics[width=3.5cm]{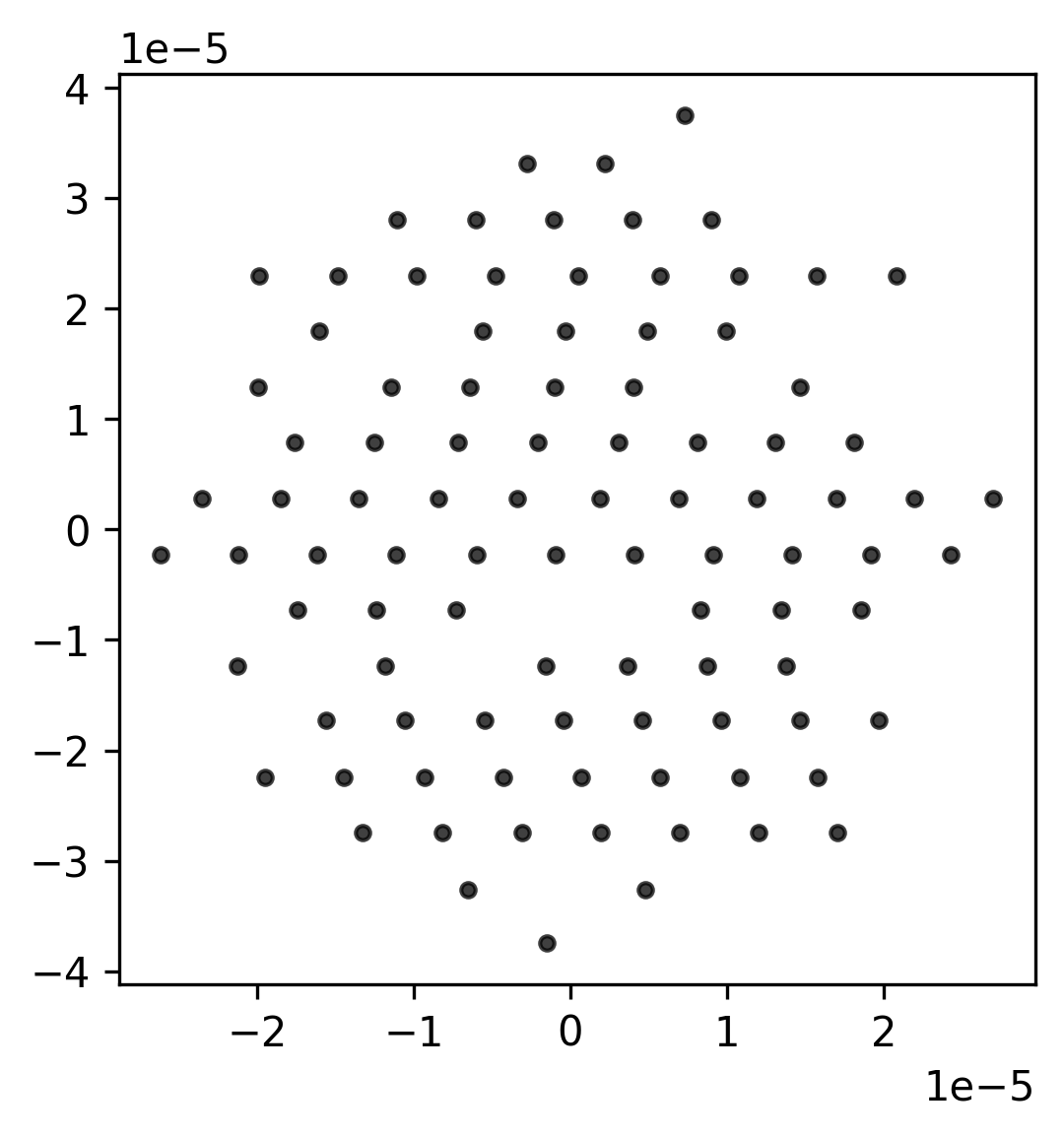}\\
	\includegraphics[width=3.5cm]{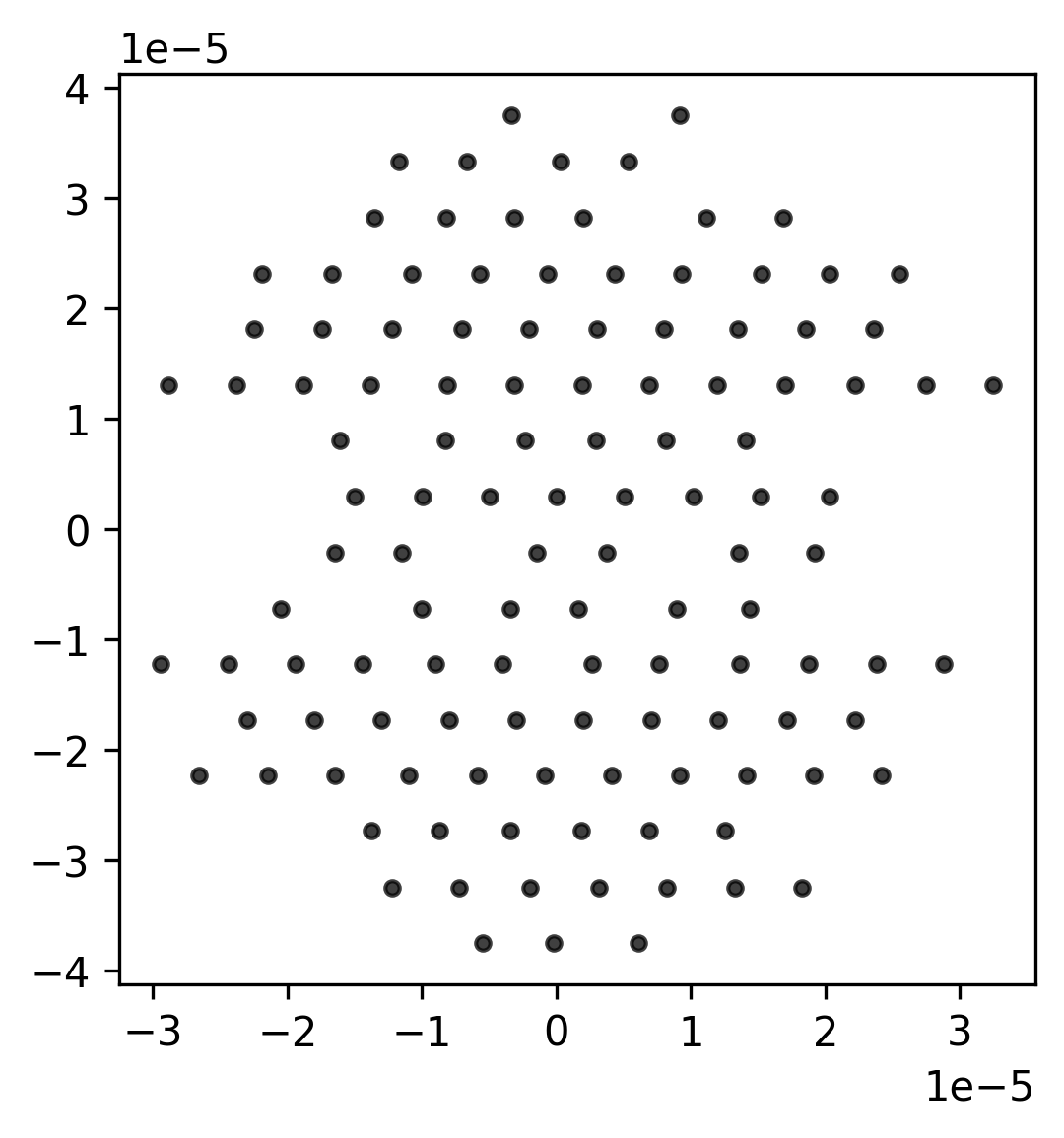}
	\includegraphics[width=3.5cm]{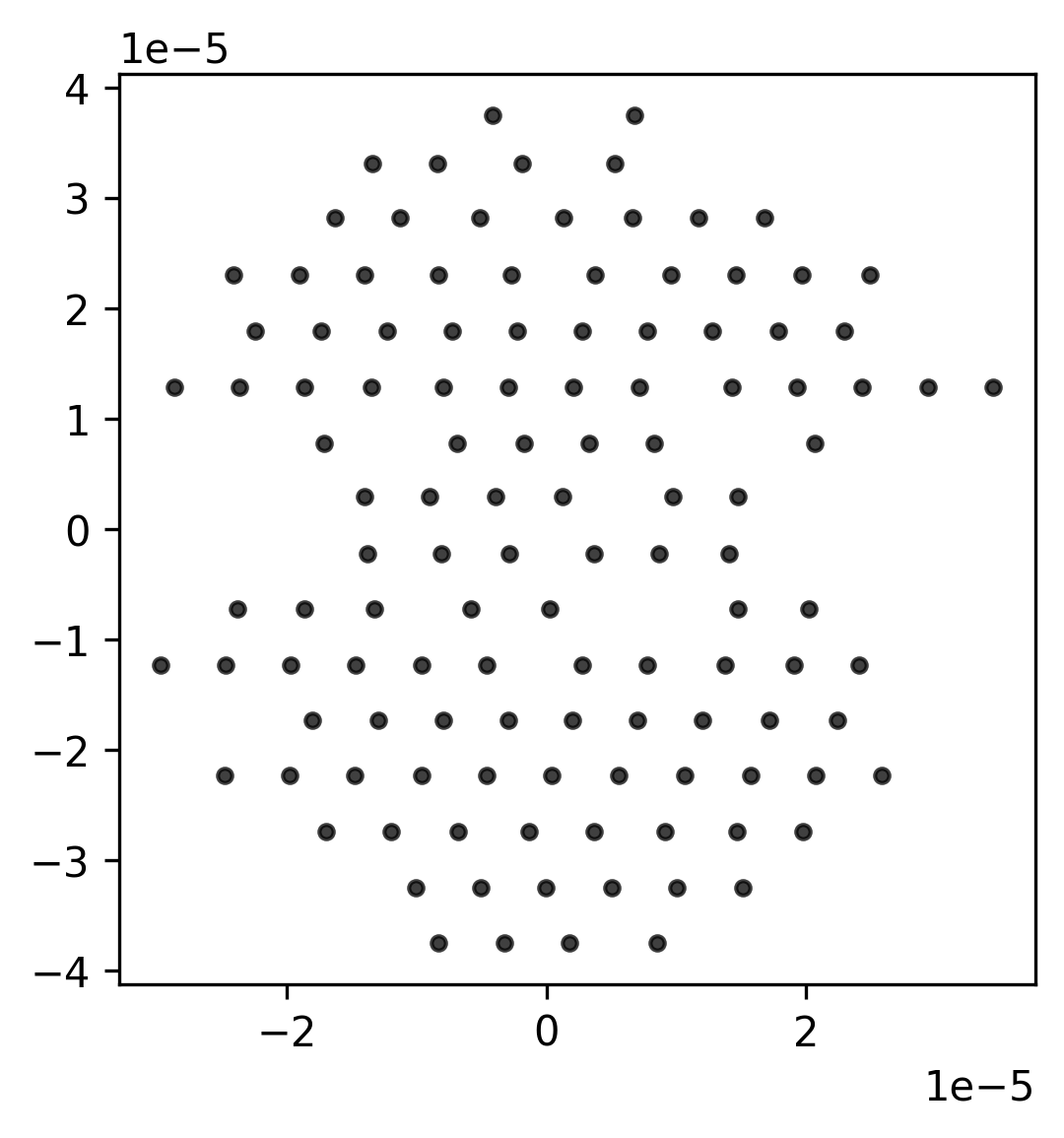}
	\includegraphics[width=3.5cm]{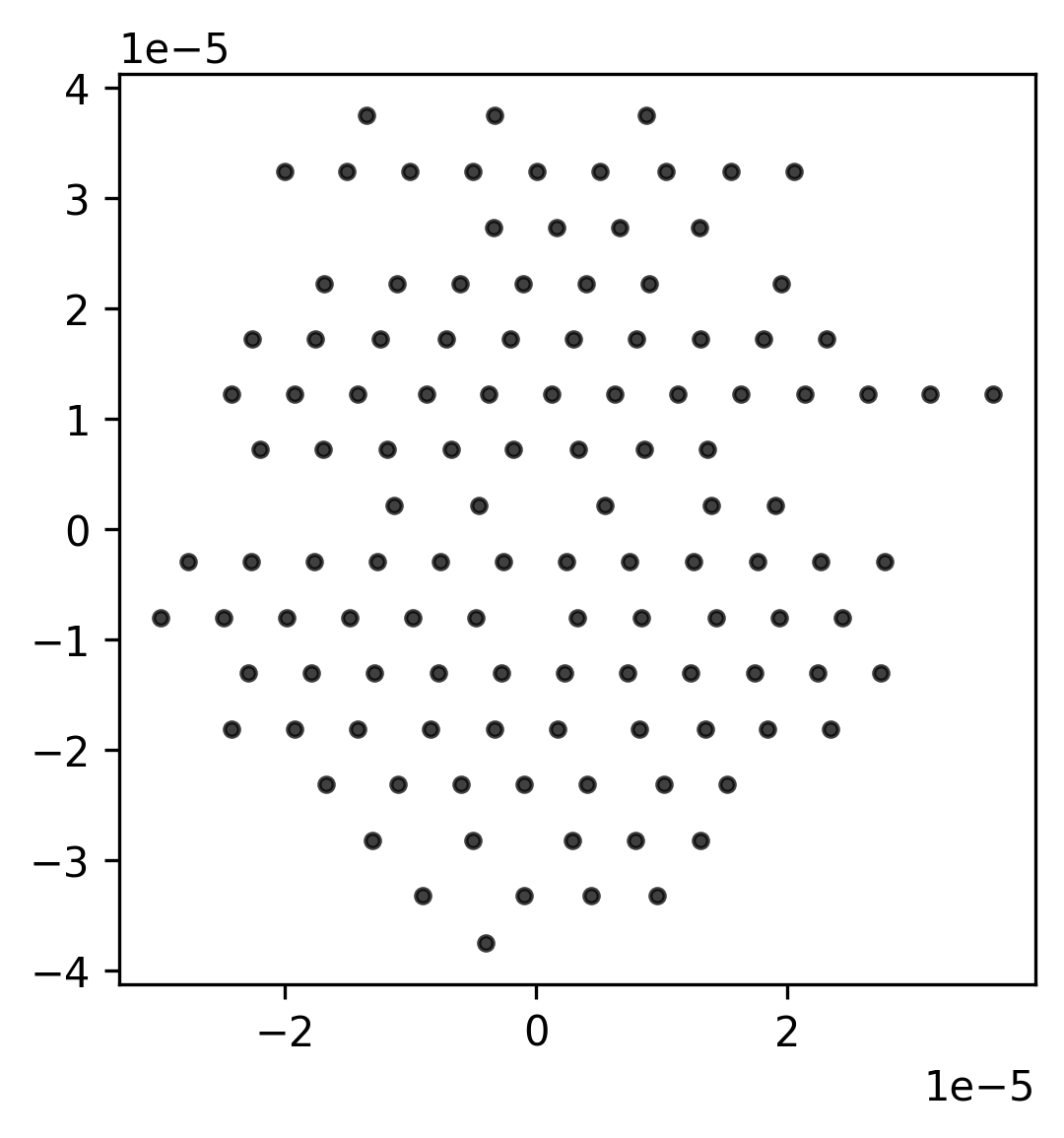}
	\includegraphics[width=3.5cm]{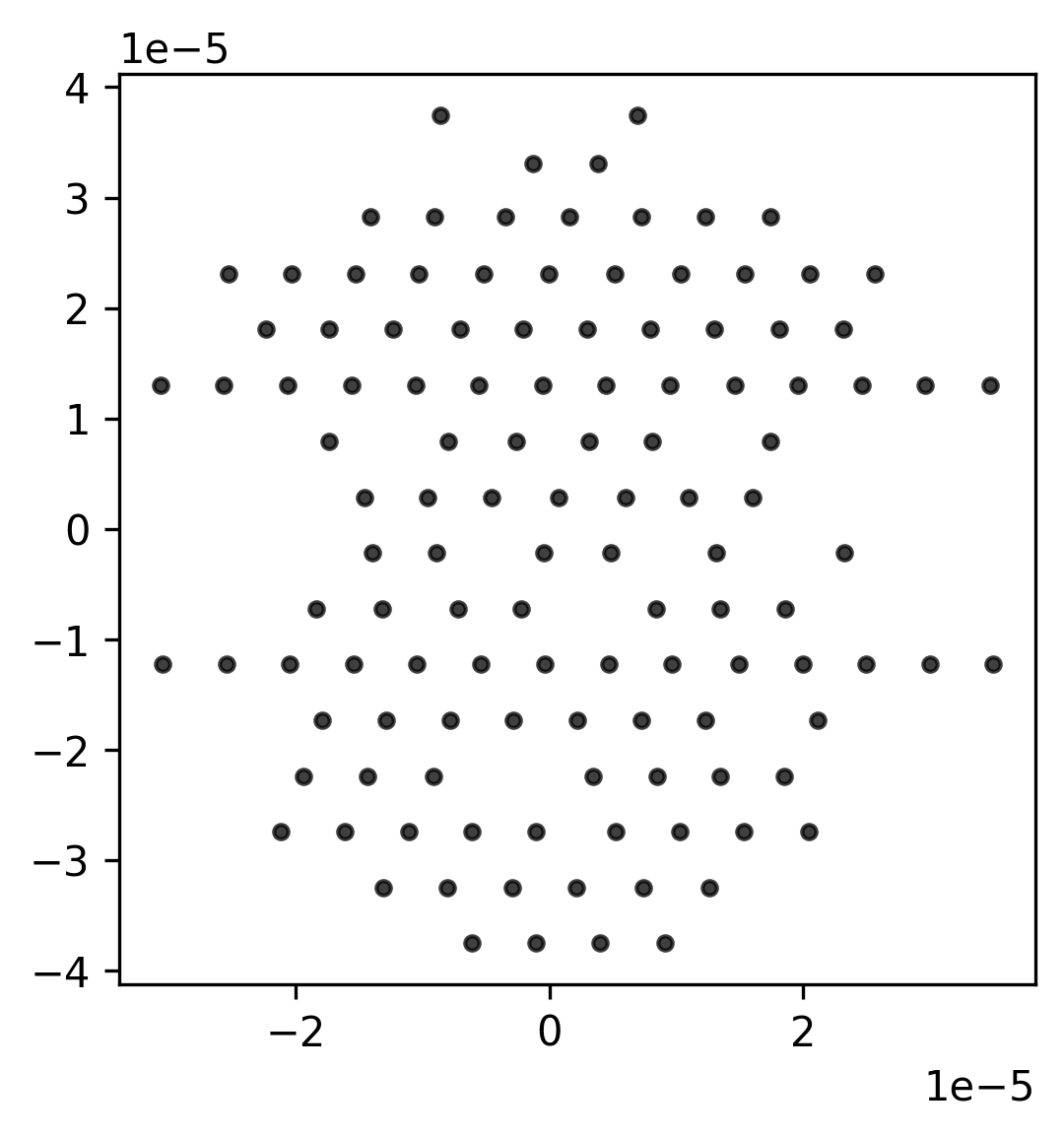}
	\includegraphics[width=3.5cm]{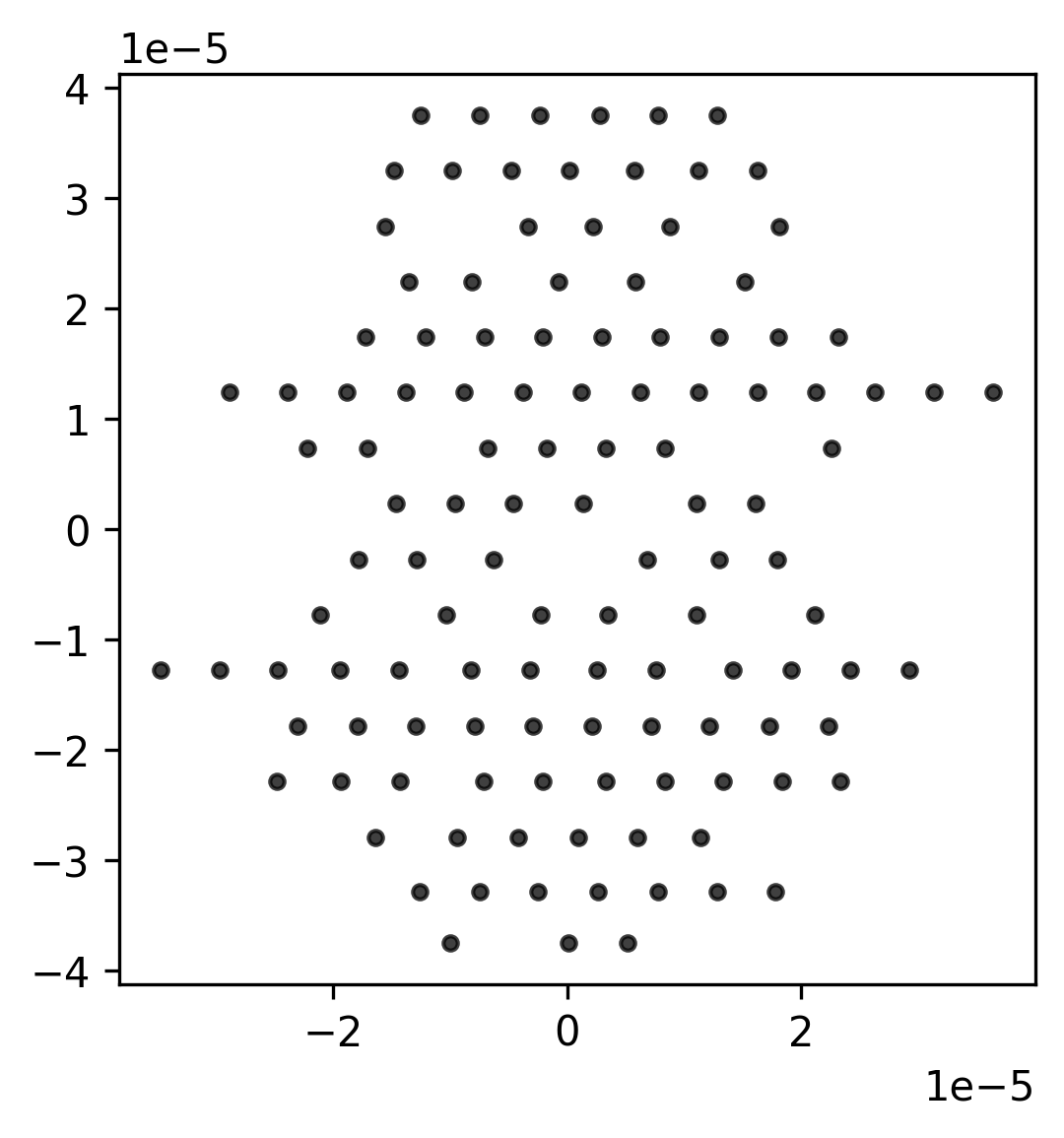}\\
	\includegraphics[width=3.5cm]{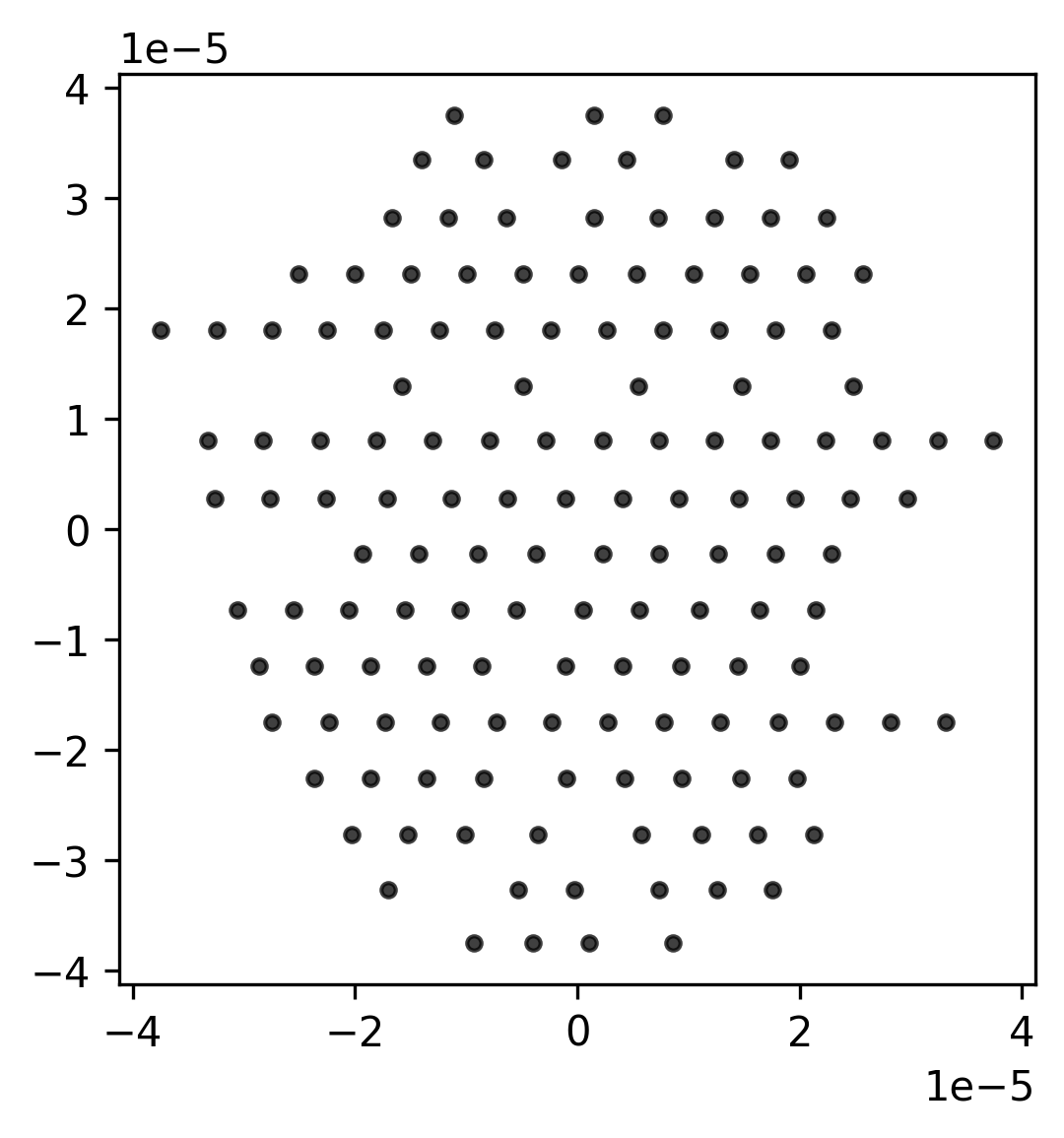}
	\includegraphics[width=3.5cm]{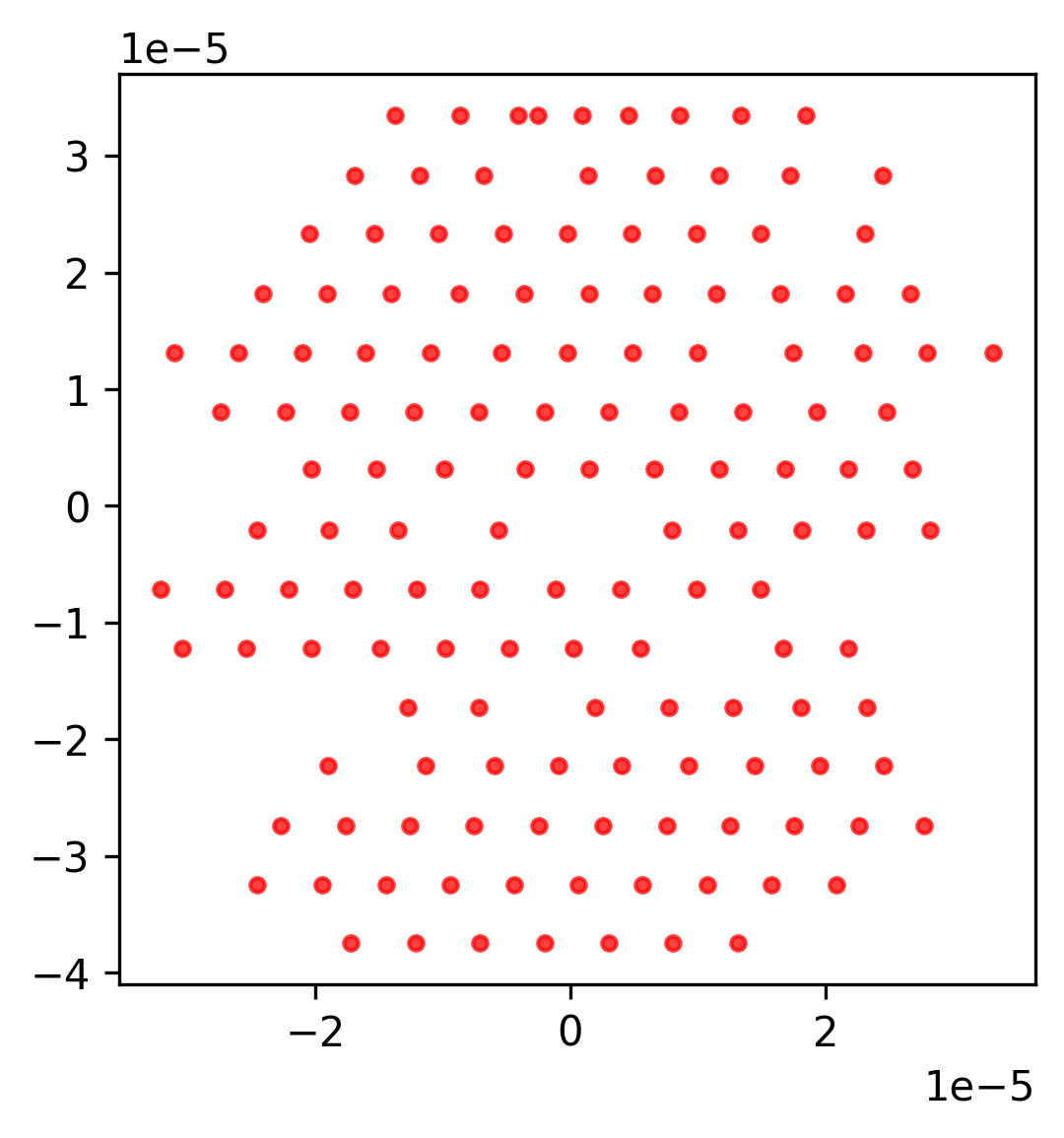}
	\includegraphics[width=3.5cm]{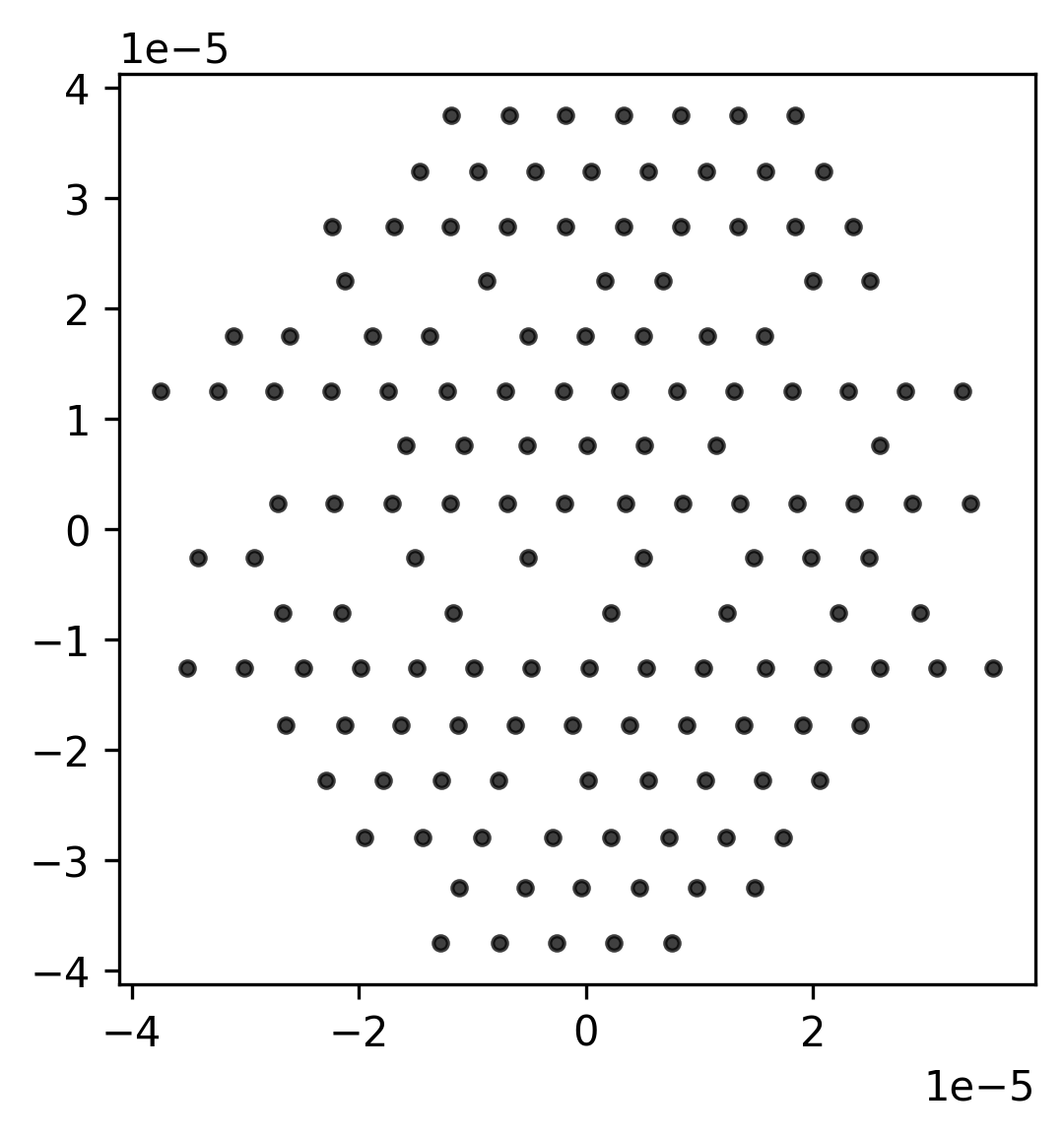}
	\includegraphics[width=3.5cm]{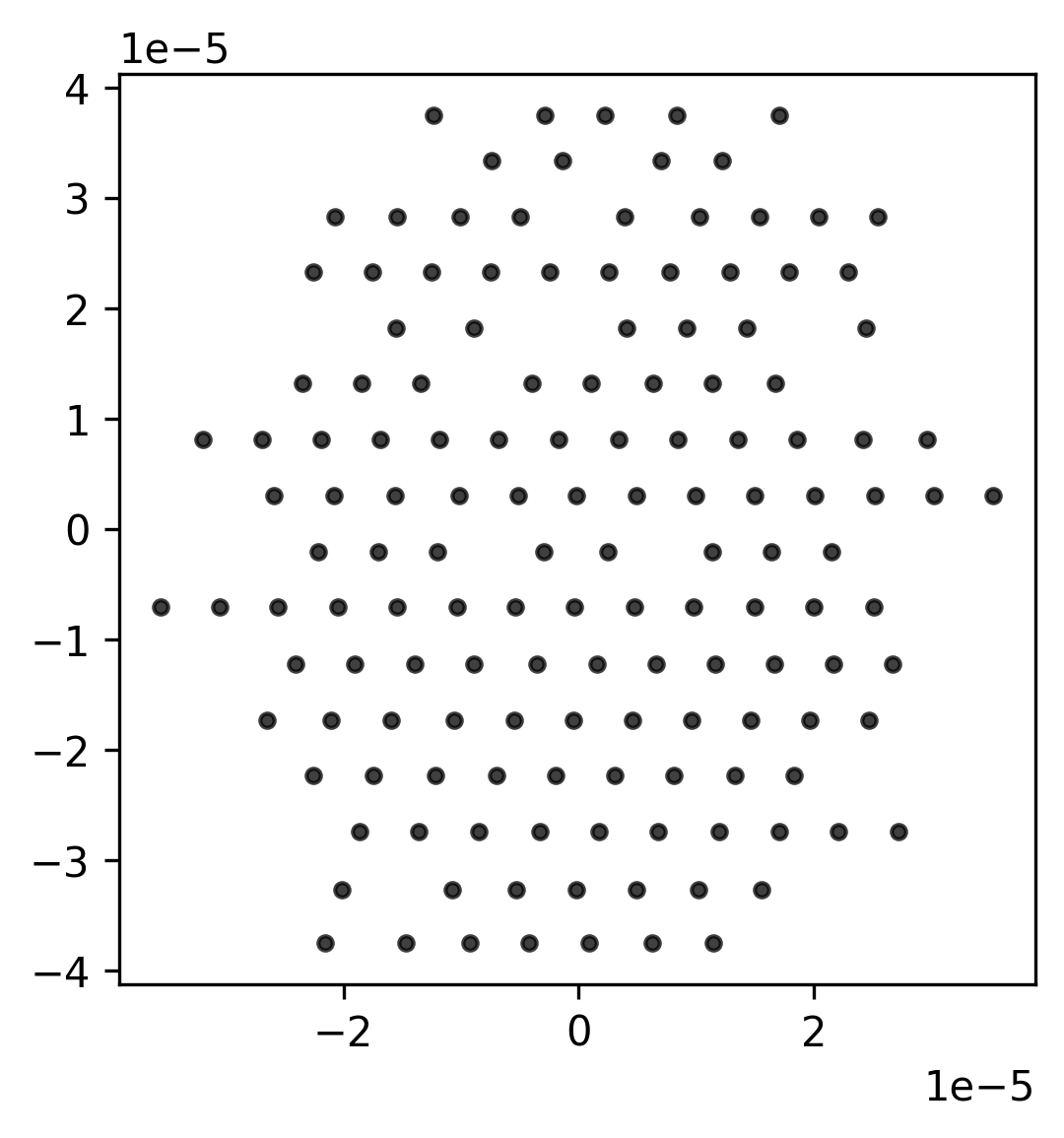}
	\includegraphics[width=3.5cm]{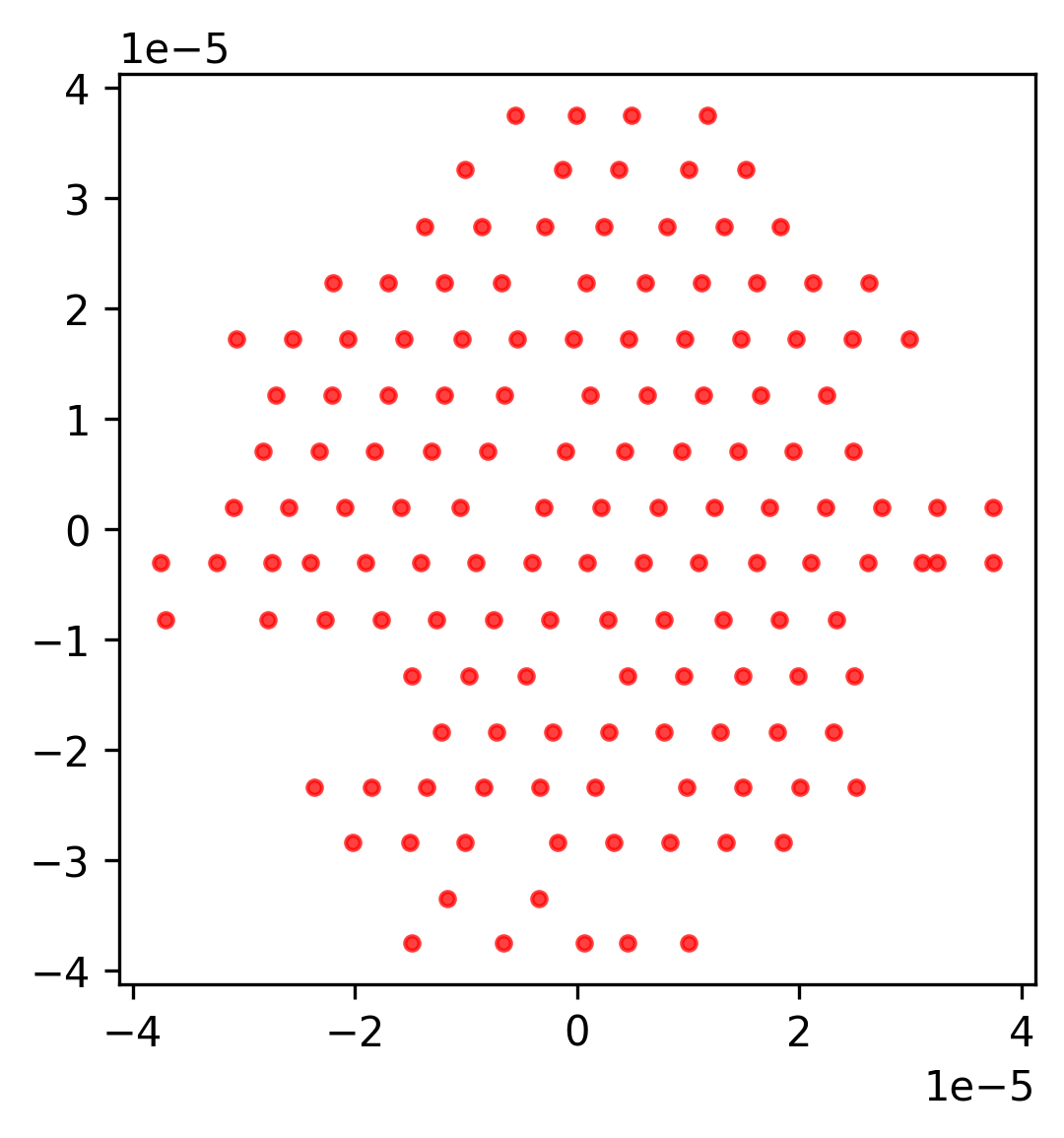}
	\caption{Atom placements computed for the encodings listed in \cref{tab:enc}. From top to bottom: binarysplit(8), stacked(16), stacked(20), stacked(24). Columns represent results for \gls{QUBO} matrices which are learned on a different subset of the data. Each plot corresponds to the output of \cref{alg:forces} for $10^5$ iterations and with stepsize $\eta=10^{-10}$. All axis are in \unit{\micro\meter}. Red placements indicate that one of the constraints was not satisfied. In both cases, it is easy to see that at least one pair of atoms violates the lower bound on the Euclidean distance between atoms.}
	\label{tab:layouts}
\end{figure*}

Atom placements for \glspl{QUBO} computed by our algorithm for the encodings from \cref{tab:enc} are shown in \cref{tab:layouts}. Therein, only 2 of 20 runs failed to satisfy all constraints with using our suggested default parameters. In fact, slightly increasing the lower bound on the Euclidean distance results in placements with a slightly larger spacing between atoms which in-turn allows the algorithm to satisfy the constraints on all cases. However, enforcing a larger spacing implies that less atoms will reside inside each others Rydberg blockade radius. Thus, reducing the actual entanglement between qubits. We thus opted for the version which allows for a higher degree of entanglement. We observe that in both scenarios with violated constraints, only single pair of atoms causes the failure. 

Finally, we note that after running \cref{alg:forces}, we recommend to pin y-coordinates which are $\epsilon$-close to each other to the exact same value, in order to mitigate numerical jitter in the placements. 
An exemplary graph that is deduced from a force-directed placement is shown in \cref{fig:graph}. 

\begin{figure}
	\centering
	\includegraphics[width=\columnwidth]{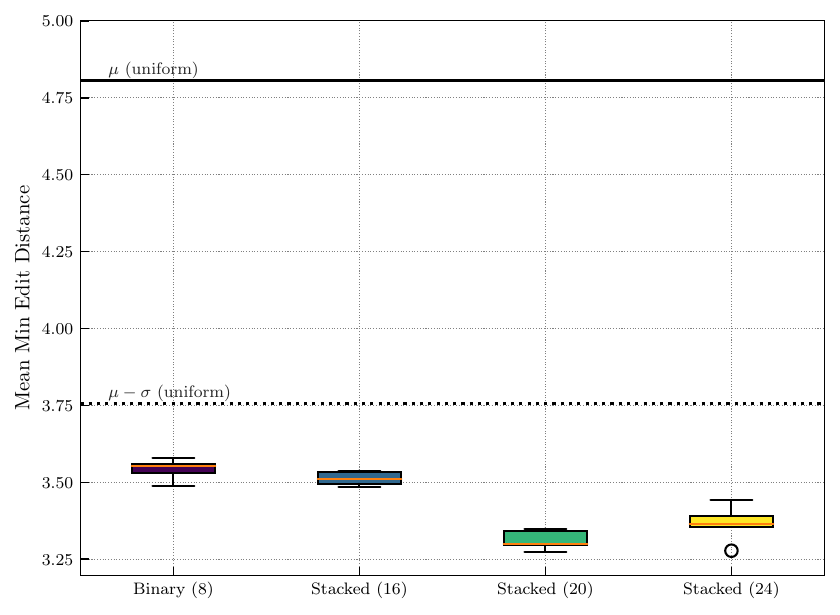}
	\caption{Box plot for the mean $\MinEditDist$ over all splits for the four encoding methods listed in \cref{tab:enc}.
	For comparison, the mean $\MinEditDist$ and one standard deviation for tokens sampled uniformly from $\cT$ is shown.}
	\label{tab:editdist}
\end{figure}

\begin{table*}
	\centering
	\caption{Random passwords from the \RockYou data set, compared to passwords sampled from Boltzmann distributions of trained \gls{QUBO} instances. The last three columns contain passwords generated with the QuEra \Aquila 256-qubit neutral atom quantum computer. The Required \gls{UD-MIS} instances are generated from \glspl{QUBO} via our Fruchterman-Reingold placement.}
\begin{tabular}{c|ccc|ccc}
	\toprule
	\RockYou  & Stacked(16) & Stacked(20) & Stacked(24) & QC Stacked(16) & QC Stacked(20) & QC Stacked(24)\\
	\midrule
	\verb|1120ha|   &\verb|akmap33|    &\verb|mes25806| &\verb|Gp9716oth| &\verb|Tunas200992| & \verb|at88gacjoE| & \verb|101mytyugar| \\
	\verb|RENAN23|     &\verb|an19in6zar| &\verb|tjnq0845| &\verb|8h4kit| &\verb|f0656ay44ey| & \verb|82therPicap| & \verb|ol77z200F44| \\
	\verb|44463811|   &\verb|arera28|    &\verb|ninycali| &\verb|cwaser24| &\verb|x5679demardo| & \verb|3Qbabff04my| & \verb|on0644s78or| \\
	\verb|r0semerran|   &\verb|onamthan?| &\verb|73atoarL|  &\verb|marsheolac08| &\verb|omagtz7866| & \verb|GTU07adom| & \verb|ooAlsarik| \\
	\verb|11691169|   &\verb|1d04f|     &\verb|812l9ol|   &\verb|bab11h94n| &\verb|idun123nejory| & \verb|35kaankydeap| & \verb|gX64erquey| \\
	\verb|daiboi1993|  &\verb|agvemD 2|  &\verb|P11rxanq|  &\verb|far12e17| &\verb|56ilthun45E| & \verb|7QumTz99| & \verb|w30g239917| \\
	\verb|1pryor|    &\verb|Hdloah6|   &\verb|22izie|    &\verb|akbipe$| &\verb|teedem28iglove| & \verb|27noNus9695| & \verb|loveul6494er28| \\
	\verb|vivicantik|    &\verb|RR2sty|    &\verb|psealotel1| &\verb|koodod04| &\verb|z20054apbe+| & \verb|7398il0ildo| & \verb|eyka3423mo14| \\
	\verb|crouite1|   &\verb|d28ku26|   &\verb|duluty07|  &\verb|09293r| &\verb|S7320ilo0654| & \verb|icZmanosryhe| & \verb|GtV123ildo| \\
	\verb|kuty06|  &\verb|usas*az|   &\verb|hent44|    &\verb|nacer19910u| &\verb|Ystevugmarda| & \verb|zEipagizse| & \verb|or73ff65marat| \\
	\bottomrule
\end{tabular}
\label{tab:gpws}
\end{table*}

\section{Experiments}
In the following, we describe our numerical password generation experiments.

\subsection{Data Set}
The \RockYou password dataset\footnote{\url{https://www.kaggle.com/datasets/wjburns/common-password-list-rockyoutxt}}~\cite{rockyou2021} is a collection of over 32 million passwords that were exposed in a data breach of the social networking application \RockYou in 2009. This dataset is widely used in cybersecurity research and password security studies, as it provides insights into user password choices, common patterns, and vulnerabilities. The dataset highlights the importance of strong password creation and the risks associated with using easily guessable passwords. Here, we use a simplified version of \RockYou which only includes printable ASCII characters. Based on these passwords, we build a tokenization $\tau$ for $T=256$ tokens via a modified Byte Pair Encoding~\cite{SennrichHB16a} following~\cite{BiesnerCGSK21}. 

\subsection{Metrics}

It is not straightforward to quantify whether a given password is ``natural'' for a human, which makes the evaluation of password generators challenging.

A common strategy is to split the data set $\cD$ into a training set $\cD^{\text{train}}$ and an evaluation set $\cD^{\text{eval}}$ \cite{BiesnerCGSK21}. 
We then build our model on $\cD^{\text{train}}$, we generate a fixed number of random passwords $\tilde{P}_c$ from it, and report the proportion of passwords in $\cD^{\text{eval}}$ that were ``guessed'' by our model, \eg, $S_{\text{overlap}}={\abs{\cD^{\text{eval}}\cap\tilde{P}_c}}/{\abs{\tilde{P}_c}}$. However, when the number of generated passwords is too small, the probability to achieve non-empty intersection with the hold-out set vanishes.  

A more informative approach for comparing strings is the \emph{edit distance} (or \emph{Levenshtein distance}) $\EditDist(\bw,\bw')$, defined as the smallest number of insertion, deletion and substitution operations needed to turn string $\bw$ into another string $\bw'$.
Given our ground-truth evaluation set $\cD^{\text{eval}}$, we can compute the minimal edit distance between a generated password $\bw_c$ and all passwords in our data set, giving us the distance to the ``most similar'' human-generated password: \begin{equation}
	\MinEditDist(\bw_c,\cD^{\text{eval}})=\min_{\bw_c'\in \cD^{\text{eval}}}~\EditDist(\bw_c,\bw_c').
\end{equation}
Computing $\EditDist$ has a runtime complexity of $\mathcal{O}(mn)$, where $m$ and $n$ are the lengths of the input strings.
Therefore, computing $\MinEditDist$ for a very large multi-set $\cD^{\text{eval}}$ quickly becomes unreasonably slow.
To address this problem, we construct a BK-tree over the strings in $\cD^{\text{eval}}$, which is a special tree structure with nodes $V=\lbrace w_c\in \cD^{\text{eval}}\rbrace$ and directed edges $E\subset V\times V$ with weights $w:E\rightarrow\BN_0$ such that \begin{enumerate*}[label=(\roman*)]
	\item all outgoing edges of a node have distinct weights, and
	\item if $e=(u,v)\in E$, then $w(e)=\EditDist(u,v)$, and for all descendants $v'$ of $v$ we also have $\EditDist(u,v')=w(e)$~\cite{burkhard.keller.1973a}.
\end{enumerate*}
This data structure reduces the number of lookups significantly, especially when the $\EditDist$ is small\footnote{The worst-case number of lookups is still $\mathcal{O}(\abs{\cD^{\text{eval}}})$ when the edit distance is large, see \url{https://github.com/benhoyt/pybktree/issues/5} (last accessed April 15, 2025)}.

\subsection{Generating Passwords}

We perform experiments to investigate the quality of passwords sampled from the Boltzmann distribution of \gls{QUBO} instances.
In particular we are interested in the quality differences depending on the chosen token encoding $\kappa$.
Based on the tokenized \RockYou password data set, we build a smaller data set $\cD^{\leq 6}$ containing only passwords consisting of at most 6 tokens (not requiring a final \EOSToken) and pad shorter passwords with \EOSToken to a length of exactly 6 tokens.

For the encodings $\kappa$ we choose binary encoding and three different stacked one-hot encodings with a different number of bits, as we want to observe the effect of using more or fewer bits (and more or fewer trainable parameters per token) on the password quality.
The exact encodings are listed in \cref{tab:enc}.
To get a better impression of our method's average behavior, we perform 5-fold cross validation by splitting $\cD^{\leq 6}$ into five subsets, running the training procedure on four subsets and using the fifth for evaluation (\ie, as a holdout set for checking if our model can generate passwords from this set, and for computing the minimum edit distance).

We based our code on the Python package \verb|qubolite|\footnote{\url{https://github.com/smuecke/qubolite} (last accessed April 15, 2025)}, which provides functionality for defining and manipulating \gls{QUBO} instances, as well as Gibbs sampling from their corresponding Boltzmann distribution.
For training, we followed this procedure for every encoding and cross validation split: \begin{enumerate}
	\item Compute the empirical distribution of binary vectors in the training set
	\item Set $t\gets 0$ and initialize a \gls{QUBO} instance with parameter matrix $\bQ^t$ depending on the token encoding (see below)
	\item Sample $10^4$ times from the Boltzmann distribution of $\bQ^t$ (see \cref{def:boltzmann}) using Gibbs sampling with a burn-in phase of 100 samples, and where only every 10-th sample is kept to reduce time dependency~\cite{Piatkowski/2018a}
	\item Compute the empirical distribution of binary vectors in the Boltzmann sample
	\item Compute the approximate gradient matrix $\hat{\bU}^t$ \cref{eq:gradapprox}
	\item Update $\bQ^{t+1}$ via ADAM~\cite{KingmaB14}
\end{enumerate}
Finally, we use $\bQ^{1000}$ as our trained \gls{QUBO} instance, from which we again draw 1000 samples with the same Gibbs sampler settings as before.
We convert the samples to passwords by de-binarizing, de-tokenizing, and stripping away trailing \EOSToken tokens.
The initial \gls{QUBO} parameter matrix $\bQ^0$ has mostly $0$-entries, except for indices $i\neq j$ that belong to the same one-hot encoding, for which we set $Q^0_{ij}=0.1$ as a small penalty, ensuring that the initial Boltzmann distribution favors valid one-hot vectors.
If a one-hot encoding produced by our model is still invalid, we perform another sampling step as a repair strategy:
If the one-hot vector contains more than one $1$, we choose one of them uniformly at random.
If the vector contains no $1$ but only zeros, we choose one index uniformly at random over all possible indices.
Using this procedure, we ensure that every model output can be interpreted as a valid password.
Mean MEDs and exemplary passwords are shown in \cref{tab:editdist} and \cref{tab:gpws}, respectively.

\section{Conclusion}

We presented the first work on generating human-like passwords from adiabatic quantum computers. 
To this end, we investigated multiple encodings of passwords into tokens and eventually, binary strings. 
The weights of a \gls{QUBO}'s Boltzmann a distribution are then trained to represent a variational distribution over passwords. 
Our empirical results based on the \RockYou dataset show that the learned models 
achieve mean $\MinEditDist$ scores which are more than one standard deviation closer to human-generated passwords compared to uniformly random token sequences. 
Qualitatively, the generated passwords exhibit patterns which are reminiscent of human generated passwords. 
This suggests that the learned \glspl{QUBO} capture important structural properties of token sequences and can hence be 
considered as a basic learned language model. 
Our results can thus be interpreted as one of the first generative \gls{NLP}-inspired models for a real-world task that runs on actual quantum computers.

Moreover, our novel force-directed 
atom placement algorithm allows us to convert learned \gls{QUBO} matrices to related \gls{UD-MIS} problem instances. This contribution is significant
beyond password generation, since 
the current generation of neutral atom quantum computers still requires the user to cope with various 
physical quantities, even if a purely combinatorial problem should be solved. Hence, our contributions provide a 
first approach for automatically generating problem specific atom placements and thus, increasing the usability of such devices. 
Future work shall investigate theoretical properties of the proposed placement algorithm, \eg, estimating the probability that all constraints are satisfied or the number of iterations until the placement has stabilized. 

Moreover, further studies may explore alternative quantum architectures for password generation, such as gate-based quantum computers. These devices can, in principle, encode and sample arbitrary probability distributions, including those derived from data-driven patterns~\cite{heese2024}. The main practical challenge is to design efficient encodings that respect the resource constraints of current quantum hardware~\cite{dasgupta2022loadingprobabilitydistributionsquantum}.